\documentclass[aps,prb,a4paper,twocolumn,showpacs,floatfix,citeautoscript,superscriptaddress]{revtex4-1}
\usepackage{graphicx}
\usepackage{amsfonts}
\usepackage{amsmath}
\usepackage{amssymb}
\usepackage{bm}

\usepackage{bbold}
\usepackage{mathtools}
\usepackage{dcolumn}
\usepackage{hyperref}
\usepackage[usenames]{color}
\usepackage{subcaption}

\usepackage{physics}

\usepackage{ragged2e}
\DeclareCaptionJustification{justified}{\justifying}
\def\beq{\begin{equation}}
\def\eeq{\end{equation}}

\newcommand{\out}[1]{{}}

\newcommand\varpm{\mathbin{\vcenter{\hbox{%
  \oalign{\hfil$\scriptstyle+$\hfil\cr
          \noalign{\kern-.3ex}
          $\scriptscriptstyle({-})$\cr}%
}}}}

\begin{document}

\title{Orbital and electronic entanglement in quantum teleportation schemes}
\author{Anna Galler}
\address{Centre de Physique Th\'eorique, Ecole Polytechnique, Institut Polytechnique de Paris, 91128 Palaiseau Cedex, France}
\author{Patrik Thunstr\"om}
\address{Department of Physics and Astronomy, Materials Theory, Uppsala University, 75120 Uppsala, Sweden}

\begin{abstract} 
With progress towards more compact quantum computing architectures, fundamental questions regarding the entanglement of indistinguishable particles need to be addressed. In a solid state device, this quest is naturally connected to the quantum correlations of electrons.
Here, we investigate the entanglement between electrons, focusing on the entanglement of modes, the entanglement of particles and the effect of particle-number superselection rules. We elucidate the formation of mode and particle entanglement in strongly correlated materials and  show that both represent important resources  in quantum information tasks such as quantum teleportation. To this end, we qualitatively and quantitatively analyze the entanglement in three electronic teleportation schemes: (i) quantum teleportation within a molecule on graphene, (ii) a nitrogen-vacancy center and (iii) a quantum dot array. 
\end{abstract}

\maketitle

\section{Introduction}
Entanglement lies at the heart of quantum mechanics and has been investigated extensively during the last decades mainly due to its importance in quantum information, cryptography and teleportation.\cite{ent_review_2009} The vast majority of studies focuses on the entanglement of distinguishable particles, while the entanglement of identical particles such as electrons has received much less attention so far. The experimental realizations of these quantum information processes consist, nevertheless, mainly of identical particles like photons \cite{Northup_photonics_2014,Volz_prl_2006}, ultracold atoms in an optical trap \cite{Leibfried_review_2003,Haffner_2008}, or electrons in a quantum dot \cite{Eriksson_2004_qdot_exp,Schaibley_PRL_2013}. The particles are instead made distinguishable by restricting their states to non-overlapping sections of the Hilbert space, for example by a macroscopical separation of their positions. However, if one aims at building a compact quantum computer, entanglement between identical particles in overlapping orbitals can no longer be neglected. There are in addition several outstanding questions regarding the entanglement between electrons in strongly correlated materials. For example, how is entanglement between the electrons formed within a material, and how does it affect the properties of the material? The purpose of the present study is to investigate the entanglement of electrons in a material and in electronic quantum information processes. To this aim we will propose and analyze three solid-state quantum teleportation protocols.

In contrast to the single definition of entanglement between distinguishable particles, there are two complementary forms of entanglement for electrons; mode entanglement\cite{zanardi_fermionic_latt,wiseman_prl_2003,barnum_prl_2004_subsystem_ind,banuls_mode_ent_2007,Marzolino_Ann_2010, friis_mode_ent,balach_prl_2013_unif,dasenbrook2016single,Amico_review_2008,marzolino_teleportation}  and particle entanglement\cite{Amico_review_2008,schliemann_slater_2001,eckert_slater_rank,Li_PRA_2001_ent,Ghirardi2002_long,Ghirardi_PRA_2004,Ghirardi2005}. The first form is based on a bipartition of the orbitals in the system, much like the bipartition used for distinguishable states, and quantifies the quantum correlations between the two sets of orbitals. It can be seen as a resource for sending quantum information between the orbital partitions. The definition of particle entanglement focuses instead directly on the quantum correlations between the electrons, {i.e.} how far the state of the system is from a statistical mixture of Fock states (single Slater determinants). As detailed in Section \ref{sec:intro_entanglement}, particle entanglement arises exclusively from particle interactions, including the interaction with a detector, while mode entanglement is formed from both non-local interactions and the non-local one-particle potentials in the Hamiltonian. 

The definition of mode entanglement can be supplemented with system and quantum protocol dependent superselection rules \cite{wiseman_prl_2003,ariano2014fermionic,banuls_mode_ent_2007,friis_mode_ent,marzolino_teleportation}. In the context of a quantum teleportation protocol, which transfers the state of an electron in a local orbital partition to an electron in a remote orbital partition,  it is natural to impose that only operations that conserve the local particle number (N-SSR) are allowed. In the following we will therefore analyze both the standard definition of mode entanglement, put forward in Ref.~\onlinecite{zanardi_fermionic_latt}, and the N-SSR restricted mode entanglement of Wiseman and Vaccaro in Ref.~\onlinecite{wiseman_prl_2003}. 

After an in depth introduction of mode and particle entanglement in Sec.~\ref{sec:intro_entanglement} and a discussion of the processes that generate mode and particle entanglement in materials in Sec.~\ref{sec:materials}, we present and analyze three examples of quantum teleportation of electrons in Sec.~\ref{sec:examples}. We find, in agreement with the developed theory, that mode and particle entanglement represent distinct resources for possible quantum information processes with identical particles. Our work provides a new perspective to the investigation of electronic entanglement and teleportation schemes (for related work see Refs.~\onlinecite{marzolino_teleportation, beenakker_fermi_sea, sauret_spin_tele, Gigena2017,  Friis_teleportation_2020, Olofsson_tele_2020}), giving concrete solid-state examples and highlighting the connection between particle entanglement and N-SSR restricted mode entanglement.  

\section{Particle and mode entanglement}
\label{sec:intro_entanglement}
The second quantization formalism offers a natural way to address states of identical particles. It is based on the notion of creation ($\hat{c}_i^\dagger$) and annihilation ($\hat{c}_i$) operators, which create and destroy an electron in the spin-orbital $i$, respectively. In the following we will only consider orthonormalized orbitals, to avoid the additional algebra associated with overlap matrices. 

All pure many-body states can be formed by applying the creation operators to the vacuum state ($\ket{0}$), which is annihilated by any annihilation operator, $\hat{c}_i \ket{0} \equiv 0$. A pure $N$-electron state\footnote{We will not consider anti-particles (positrons) in this study which implies that any state vector must have a fixed number of electrons} can hence be written
\begin{equation}
\ket{\psi} = \sum_{\mathbf{i} \in \mathcal{S}_N} A_{\mathbf{i}} \hat{S}^\dagger_{\mathbf{i}} \ket{0}\label{eqn:psi}
\end{equation}
where the Slater determinant index $\mathbf{i} \in \mathcal{S}_N$ is an ordered sequence of $N$ orbital indices, {i.e.} $\mathbf{i}_1 < \mathbf{i}_2 < \ldots < \mathbf{i}_N$, and $\hat{S}^\dagger_{\mathbf{i}} \equiv \hat{c}^\dagger_{\mathbf{i}_1}\hat{c}^\dagger_{\mathbf{i}_2} \cdots \hat{c}^\dagger_{\mathbf{i}_N}$. A state that can be written as $\hat{S}^\dagger_{\mathbf{i}} \ket{0}$ in a given orbital basis is known as a Slater determinant. In the following we will call a state that is a Slater determinant in some orbital basis, but not necessarily the given one, a Fock state. The reason for this distinction will become clear when we define particle and mode entanglement.

The choice of orbital basis does not hold any physical significance in itself. A change of orbital basis $\hat{c}_i^\dagger \rightarrow \hat{c}_i^{\prime\dagger}$, where
\begin{equation}
\hat{c}_j^{\prime\dagger} \equiv \sum_{i} \hat{c}_i^\dagger V_{ij}^\dagger,\label{eqn:cprime} 
\end{equation}
and $V$ is a unitary transformation, can be performed by substituting the identity
\begin{equation}
\hat{c}_i^\dagger = \sum_{j} \hat{c}_j^{\prime\dagger} V_{ji}
\end{equation}
into Eq.~(\ref{eqn:psi}). The fermionic commutation relation $\hat{c}^\dagger_{i}\hat{c}^\dagger_{j} = -\hat{c}^\dagger_{j}\hat{c}^\dagger_{i}$ can then be used to sort the creation operators according to the selected orbital order. For example, the state 
\begin{equation}
\ket{\psi'}=\frac{1}{2}(\hat{c}_{1\uparrow}^\dagger+\hat{c}_{1\downarrow}^\dagger)(\hat{c}_{2\uparrow}^\dagger+\hat{c}_{2\downarrow}^\dagger)\ket{0},\label{eqn:psiexample}
\end{equation}
represents two electrons located in the spin-orbitals 1$\uparrow$, 1$\downarrow$, 2$\uparrow$, and 2$\downarrow$. In the orbital basis 
\begin{align}\label{eq:trafo_sep}
\hat{c}_{1\uparrow x}^\dagger&=\frac{1}{\sqrt{2}}(\hat{c}_{1\uparrow}^\dagger+\hat{c}_{1\downarrow}^\dagger) \hspace{3em} \hat{c}_{2\uparrow x}^\dagger=\frac{1}{\sqrt{2}}(\hat{c}_{2\uparrow}^\dagger+\hat{c}_{2\downarrow}^\dagger)   \nonumber \\
\hat{c}_{1\downarrow x}^\dagger&=\frac{1}{\sqrt{2}}(\hat{c}_{1\uparrow}^\dagger-\hat{c}_{1\downarrow}^\dagger)
\hspace{3em} \hat{c}_{2\downarrow x}^\dagger=\frac{1}{\sqrt{2}}(\hat{c}_{2\uparrow}^\dagger-\hat{c}_{2\downarrow}^\dagger) ,
\end{align}
it takes the simple product form
\begin{equation}
\ket{\psi'}=\hat{c}_{1\uparrow x}^\dagger\hat{c}_{2\uparrow x}^\dagger\ket{0}.
\end{equation}
The unitary transformation in Eq.~(\eqref{eq:trafo_sep}) corresponds to a $\pi/2$ spin rotation around the y-axis. 

The product form of a Fock state, {i.e.} that it can be written as a single Slater determinant $\hat{S}^\dagger_{\mathbf{i}} \ket{0} = \hat{c}^\dagger_{i_1}\hat{c}^\dagger_{i_2} \cdots \hat{c}^\dagger_{i_N} \ket{0}$ in some orbital basis, gives it properties closely related to those of a product state of distinguishable particles \cite{Ghirardi2005,eckert_slater_rank}. For example, an unknown Fock state can be fully characterized by single-particle measurements of the orbital occupation. It can therefore in principle be described by a hidden-variable theory, where the hidden variables specify which orbitals are fully occupied. 
This has lead to the concept of {\em particle entanglement}, which identifies the Fock states as being non-entangled\cite{schliemann_slater_2001, Ghirardi2005,eckert_slater_rank,Ghirardi_PRA_2004,Ghirardi2005,Ghirardi2002_long,Kraus_fermion}. 
It should however be noted that it is only when the occupied orbitals of a Fock state are local that the hidden variables description become local as well. The other 'non-local' Fock states can hence still potentially be used as a resource in a quantum computational algorithm. 

Although the choice of orbital basis does not hold any physical significance, the way the state can be written in terms of creation operators still affects orbital-dependent quantities. For example, given a set of orbitals that belongs to Alice $A$, and a set that belongs to Bob $B$, one may ask whether the electrons within one orbital partition can be described independently of the electrons in the other partition. This is the defining idea behind mode entanglement \cite{zanardi_fermionic_latt}, which follows closely the concept of entanglement between distinguishable particles but applied to the orbital occupation. A state $\ket{\psi}$ is mode entangled with respect to the orbital partitions $A$ and $B$ unless it can be written as the product
\begin{equation}
\ket{\psi} = \Big( \sum_{\mathbf{i} \in \mathcal{S}^A} A_{\mathbf{i}} \hat{S}^\dagger_{\mathbf{i}} \Big) \Big( \sum_{\mathbf{j} \in \mathcal{S}^B} B_{\mathbf{j}} \hat{S}^\dagger_{\mathbf{j}} \Big) \ket{0},
\end{equation}
where $\mathbf{i} \in \mathcal{S}^A$ denotes all possible combinations of occupied orbitals in $A$. Even a single electron state can thus be mode entangled if the corresponding occupied orbital is shared between $A$ and $B$. The simplest possible mode entangled state is hence
\begin{equation} 
\ket{\psi} = \frac{1}{\sqrt{2}} \Big(\hat{c}_{1}^\dagger + \hat{c}_{2}^\dagger \Big) \ket{0},
\end{equation}
where orbital 1 and 2 belong to partition $A$ and $B$, respectivly. A less trivial example is given by $\ket{\psi'}$ in Eq.~(\ref{eqn:psiexample}), if the spin up orbitals belong to $A$ and the spin down orbitals to $B$. A third example of a mode entangled state is
\begin{equation}
\label{eq:2fer_ent} 
\ket{\psi''} = \frac{1}{\sqrt{2}} \Big(\hat{c}_{1\uparrow}^\dagger\hat{c}_{2\downarrow}^\dagger + \hat{c}_{1\downarrow}^\dagger\hat{c}_{2\uparrow}^\dagger \Big) \ket{0}.
\end{equation}
This state is special since it is mode entangled with respect to {\em any} bipartition of the spin-orbitals 1$\uparrow$, 1$\downarrow$, 2$\uparrow$, and 2$\downarrow$. Such state can not be a Fock state, since a fully occupied or empty orbital would form a separable subspace, which implies that the state must be particle entangled \cite{schliemann_slater_2001, Ghirardi2005,eckert_slater_rank,Ghirardi_PRA_2004}.  The converse, that a particle entangled state is mode entangled with respect to {\em any} bipartition is not true in general since some of the orbitals can still be fully occupied or empty in a particle entangled state. 

A quantum teleportation scheme transfers the unknown state of an electron from one local orbital partition $A$ (belonging to Alice) to another orbital partition $B$ (Bob) without a {\em phase coherent} transport of electrons or other information carriers between the two orbital partitions. Only operations that conserve the particle number of the two orbital partitions are hence allowed in the teleportation protocol.
The unknown state is in general entangled with electrons in a third remote partition, which implies that also its relative phase needs to be transferred. 
Alice and Bob need additional electrons to carry out the teleportation, but in order to have well-defined conditions for a successful teleportation these electrons should not initially be entangled with the unknown state.
In particular, a successful teleportation should not be affected by any subsequent local measurements performed by Alice. Alice may hence perform a total occupation measurement after the teleportation, without affecting the result, using the local occupation number operator
\begin{equation}
\hat{N}_A = \sum_{i \in A} \hat{c}^\dagger_i \hat{c}_i,
\end{equation}
where $i \in A$ runs over all the spin-orbitals in $A$. 
Since all the allowed operators in the teleportation protocol preserve the number of electrons in $A$, they commute with $\hat{N}_A$. This implies that Alice may perform a total occupation measurement of her orbital partition at the very start of the teleportation protocol and still get the same result on average\cite{wiseman_prl_2003}. The N-SSR restricted mode entanglement corresponds to the mode entanglement of the system but after a projective total occupation measurement of the local orbital partition $A$ or $B$.
The measurement of the occupation number projects an N-electron state $\ket{\psi_N}$ into $\hat{P}^{(n)}_A \ket{\psi_N}$ with probability $\bra{\psi_N} \hat{P}^{(n)}_A \ket{\psi_N}$, where the projection operator $\hat{P}^{(n)}_A$ is given by
\begin{equation}
\label{eq:projn}
\hat{P}^{(n)}_A = \sum_{\substack{\mathbf{i} \in \mathcal{S}^B \\ \mathbf{j} \in \mathcal{S}^A_n}} \hat{S}_{\mathbf{i}} \hat{P}_{\mathbf{j}\mathbf{i}},
\end{equation}
where the operator
\begin{eqnarray}
\hat{P}_{\mathbf{j}\mathbf{i}} & = \hat{S}_{\mathbf{j}}^\dagger \ket{0}\bra{0} \hat{S}_{\mathbf{j}} \hat{S}_{\mathbf{i}}\label{eqn:projection} 
\end{eqnarray}
has been introduced for later convenience. A pure N-electron state is therefore N-SSR mode entangled unless it can be written in the form
\begin{equation}
\ket{\psi_N} = \sum_{n=0}^{N}\Big( \sum_{\mathbf{i} \in \mathcal{S}^A_n} A_{\mathbf{i}} \hat{S}^\dagger_{\mathbf{i}} \Big) \Big( \sum_{\mathbf{j} \in \mathcal{S}^B_{N-n}} B_{\mathbf{j}} \hat{S}^\dagger_{\mathbf{j}} \Big) \ket{0}.
\end{equation}
As shown in Appendix \ref{app:nssrparticle}, if an N-electon state has N-SSR restricted mode entanglement, it will also be particle entangled after the projective occupation number measurement. N-SSR restricted mode entanglement is from this perspective a hybrid of mode entanglement and particle entanglement, where the mode entanglement gives the non-local correlation while the particle entanglement provides the handle to access the correlation.

\subsection{Entanglement measures}\label{sec:measure}
Since the definition of mode entanglement follows closely the definition of entanglement for distinguishable particles, although applied to the orbital occupation, it can be measured in a similar fashion. Given a bipartition of the orbitals into $A$ and $B$, the partial trace of $\ket{\psi}$ over the orbitals in $B$ gives the reduced density matrix
\begin{equation} 
\label{eq:red_dmatrix}
\rho_A = \Tr_B\Big[ \ket{\psi}\bra{\psi} \Big] \equiv \sum_{\substack{\mathbf{i} \in \mathcal{S}^B \\ \mathbf{j,k} \in \mathcal{S}^A}} \hat{P}_{\mathbf{j}\mathbf{i}} \ket{\psi}\bra{\psi} \hat{P}^{\dagger}_{\mathbf{i}\mathbf{k}},
\end{equation}
where the projection operator $\hat{P}_{\mathbf{j}\mathbf{i}}$ was introduced in Eq.~(\ref{eqn:projection}). The mode entanglement in $\ket{\psi}$ between $A$ and $B$ is converted to entropy in $\rho_A$ (and $\rho_B$), so the (linear) entropy $S[\rho_A] = \Tr[\rho_A(1-\rho_A)]$ serves as a measure of the mode entanglement. However, if the initial state is mixed, its non-zero entropy is partially transferred to $\rho_A$ and $\rho_B$, which implies that $S[\rho_A]$ only gives an upper bound to the mode entanglement in this case. 

N-SSR restricted mode entanglement is also measured using the entropy of the reduced density matrix, but as an average after the local occupation has been resolved. 
The N-SSR restricted mode entanglement is hence given by
\begin{equation}
S^N[\rho_A] = \sum_{n} \Tr\Big[\rho^{(n)}_A\Big] S[\frac{\rho^{(n)}_A}{\Tr\Big[\rho^{(n)}_A\Big]}],
\end{equation}
where the particle resolved reduced density matrix $\rho^{(n)}_A$ of the N-electron state $\ket{\psi}$ is given by
\begin{equation}
\label{eq:redn_dmatrix}
\rho^{(n)}_A = \sum_{\substack{\mathbf{i} \in \mathcal{S}^B_{N-n} \\ \mathbf{j,k} \in \mathcal{S}^A_n}} \hat{P}_{\mathbf{j}\mathbf{i}} \ket{\psi}\bra{\psi} \hat{P}^{\dagger}_{\mathbf{i}\mathbf{k}}.
\end{equation}

The particle entanglement of a pure state can also be measured using the entropy of a reduced density matrix \cite{Amico_review_2008}, but in this case of the one-particle reduced density matrix $\rho^{(1\mathrm{p})}_{ij} = \bra{\Psi}\hat{c}^{\dagger}_j \hat{c}_i \ket{\Psi}$. The entropy of $\rho^{(1\mathrm{p})}/\braket{\Psi}{\Psi}$,
\begin{align}
S[\ket{\Psi}] & \equiv \braket{\Psi}{\Psi} S\Big[\frac{\rho^{(1\mathrm{p})}}{\braket{\Psi}{\Psi}}\Big] \nonumber\\ & = \braket{\Psi}{\Psi} \Tr\Big[\frac{\rho^{(1\mathrm{p})}}{\braket{\Psi}{\Psi}}\Big(1-\frac{\rho^{(1\mathrm{p})}}{\braket{\Psi}{\Psi}}\Big)\Big],\label{eq:ent_entropy}
\end{align}
is zero for any Fock state, positive for a particle-entangled state, and invariant under any unitary orbital transformation $c^\dagger_i \rightarrow c^{\prime\dagger}_i$. Again, if the initial state is mixed, then $S[\rho^{(1\mathrm{p})}]$ will only give an upper bound to the particle entanglement. An alternative entanglement measure is based on the geometric distance to the closest Fock state\cite{Vedral_PRL_1997_ent_meas,patrik_entanglement_prl,zhang2014optimal},
\begin{equation}
\label{eq:ent_geometric}
E_G[\ket{\Psi}] = \bra{\Psi}\ket{\Psi}-\max\limits_{\ket{\Psi'}}\abs{\bra{\Psi'}\ket{\Psi}}^2
\end{equation}
where $\ket{\Psi'}$ is restricted to be pure and separable. If the system has only two electrons it is straightforward to show that $E_G[\ket{\Psi}] = \braket{\Psi}{\Psi} - \rho^{(1\mathrm{p})}_{\max}$, where $\rho^{(1\mathrm{p})}_{\max}$ is the largest eigenvalue of the one-particle reduced density matrix $\rho^{(1\mathrm{p})}$ \cite{patrik_entanglement_prl}. In the case of three or more electrons the search for the closest Fock state becomes much more involved, as illustrated by the state
\begin{equation}\label{eq:3ferm_ent}
\ket{\psi'''}=\frac{1}{\sqrt{3}}(\hat{c}_{1\downarrow}^\dagger\hat{c}_{2\uparrow}^\dagger\hat{c}_{3\uparrow}^\dagger-\hat{c}_{1\uparrow}^\dagger\hat{c}_{2\downarrow}^\dagger\hat{c}_{3\uparrow}^\dagger+\hat{c}_{1\uparrow}^\dagger\hat{c}_{2\uparrow}^\dagger\hat{c}_{3\downarrow}^\dagger)\ket{0},
\end{equation}
which is composed of three Slater determinants. At first glance, the maximum squared overlap of $\ket{\psi'''}$ with a Fock state seems to be 1/3. However, a parametrized search over all unitary orbital transformations\cite{patrik_entanglement_prl} yields the transformation
\begin{equation}
\hat{c}_{i\uparrow}^{\prime\dagger} = \frac{1}{\sqrt{3}}(\sqrt{2}\hat{c}_{i\uparrow}^{\dagger}+\hat{c}_{i\downarrow}^{\dagger}) 
\hspace{2em}
\hat{c}_{i\downarrow}^{\prime\dagger} = \frac{1}{\sqrt{3}}(-\hat{c}_{i\uparrow}^{\dagger}+\sqrt{2}\hat{c}_{i\downarrow}^{\dagger})
\end{equation}
that allows us to rewrite the state in Eq.~\eqref{eq:3ferm_ent} as
\begin{align}\label{eq:3ferm_ent_rot}
\ket{\psi'''} & =\frac{2}{3}\hat{c}_{1\uparrow}^{\prime\dagger}\hat{c}_{2\uparrow}^{\prime\dagger}\hat{c}_{3\uparrow}^{\prime\dagger}\ket{0} + \frac{1}{3}\hat{c}_{1\uparrow}^{\prime\dagger}\hat{c}_{2\downarrow}^{\prime\dagger}\hat{c}_{3\downarrow}^{\prime\dagger}\ket{0} + \frac{1}{3}\hat{c}_{1\downarrow}^{\prime\dagger}\hat{c}_{2\uparrow}^{\prime\dagger}\hat{c}_{3\downarrow}^{\prime\dagger}\ket{0} \nonumber \\
& - \frac{1}{3}\hat{c}_{1\downarrow}^{\prime\dagger}\hat{c}_{2\downarrow}^{\prime\dagger}\hat{c}_{3\uparrow}^{\prime\dagger}\ket{0} - \frac{\sqrt{2}}{3}\hat{c}_{1\downarrow}^{\prime\dagger}\hat{c}_{2\downarrow}^{\prime\dagger}\hat{c}_{3\downarrow}^{\prime\dagger}\ket{0}. 
\end{align}
The first Slater determinant, $\hat{c}_{1\uparrow}^{\prime\dagger}\hat{c}_{2\uparrow}^{\prime\dagger}\hat{c}_{3\uparrow}^{\prime\dagger}\ket{0}$, on the right hand side of Eq.~\eqref{eq:3ferm_ent_rot} has a weight of $2/3 > 1/\sqrt{3}$. $\ket{\psi'''}$ is hence particle-entangled with a geometric entanglement measure of $E_G=1-|2/3|^2=5/9$. The entropic entanglement measure for the same state $\ket{\psi'''}$ can be calculated from its one-particle reduced density matrix, which in the basis \{1$\uparrow'$,1$\downarrow'$,2$\uparrow'$,2$\downarrow'$,3$\uparrow'$,3$\downarrow'$\} reads   
\begin{equation}\label{eq:dmatrix_1}
\rho^{(1\mathrm{p})}= \frac{1}{9}
\begin{pmatrix}
5&-\sqrt{2}&0&0&0&0\\
-\sqrt{2}&4&0&0&0&0\\
0&0&5&\sqrt{2}&0&0\\
0&0&\sqrt{2}&4&0&0\\
0&0&0&0&5&\sqrt{2}\\
0&0&0&0&\sqrt{2}&4\\
\end{pmatrix}.
\end{equation}
By inserting $\rho^{(1\mathrm{p})}$ into Eq.~\eqref{eq:ent_entropy} we obtain an entanglement entropy of $S=4/3$. 
 For comparison, the two-fermion entangled state of Eq.~\eqref{eq:2fer_ent} has an entanglement entropy of $S=1$ and a geometric entanglement measure of $E_G=1-1/2=1/2$. 
In the result section we will use these entanglement measures to analyze the mode entanglement and the particle entanglement in three different teleportation schemes involving identical particles.

\subsection{Entanglement in a material}
\label{sec:materials}
The evolution of the electrons within a material is governed by the Schr{\"o}dinger equation and a many-body Hamiltonian composed of a one-particle term ($\hat{H}^{(1)}$) and the two-particle Coulomb interaction ($\hat{U}$). The one-particle term $\hat{H}^{(1)}$ can in general be written
\begin{equation}
\hat{H}^{(1)} = \sum_{mn} H^{(1)}_{mn} \hat{c}_{m}^\dagger \hat{c}_{n},
\end{equation}
where $H^{(1)}_{mn} = \bra{0} \hat{c}_{m} \hat{H}^{(1)} \hat{c}_{n}^\dagger \ket{0}$ is the matrix representation of $\hat{H}^{(1)}$ evaluated in the one-particle Slater determinant basis $\hat{c}_{n}^\dagger \ket{0}$. The unscreened two-particle Coulomb interaction is given by
\begin{equation}
\hat{U} = \sum_{\sigma\sigma'} \iint \hat{c}^\dagger_{r,\sigma} \hat{c}^\dagger_{r',\sigma'} \frac{1}{|{r} - {r}^{\prime}|} \hat{c}_{r',\sigma'} \hat{c}_{r,\sigma} drdr',\label{eqn:u}
\end{equation}
where $r$ and $\sigma$ are the position and the spin of the electron, respectively. 

All materials, except solid hydrogen, have some contracted atomic-like (core) orbitals that are always completely filled with electrons due to their large attractive interaction with the nucleus. The electrons in these core orbitals can therefore be traced out of the system. The interaction term between these core electrons and the remaining (valence) electrons is then reduced to an additional effective one-particle potential term in $\hat{H}^{(1)}$. The kinetic energy term in $\hat{H}^{(1)}$ cause the remaining atomic orbitals to hybridize with the orbitals of the neighbouring atoms, but the strength of the hybridization depends strongly on the overlap between the orbitals. The $3d$-orbitals of first row transition metal atoms and the $f$-orbitals of the lanthanides and actinides are particularly contracted compared to the more extended valence s- and p-orbitals. The weeker hybridization increase the relative importance of the $\hat{U}$ term within the contracted orbitals, while the strongly hybridizing valence s- and p-orbitals are often well-described by mean-field-like approximations that reduce the $\hat{U}$ term to an effective potential in $\hat{H}^{(1)}$ \cite{Anisimov1997_1,lichtenstein_dft_dmft,Held_review}. Strongly correlated materials, {i.e.} materials that can not even qualitatively be described without particle entanglement by an effective $\hat{H}^{(1)}$, have therefore in general partially filled localized $d$- or $f$-orbitals.

The two terms of the Hamiltonian, $\hat{H}^{(1)}$ and $\hat{U}$, do not in general commute, but the Trotter decomposition of the resulting evolution operator
\begin{equation}
e^{i\hat{H}t} = \lim_{M \rightarrow \infty} \Big( e^{i\hat{H}^{(1)} t/M} e^{i\hat{U} t/M} \Big)^M,
\end{equation}
allows us to consider the effect of the one-particle unitary operator $\hat{W}^{(1)} = e^{i\hat{H}^{(1)} t/M}$ and the two-particle unitary operator $\hat{W}^{(2)} = e^{i\hat{U} t/M}$ separately.

It is well-known that the evolution given by $\hat{W}^{(1)}$ simply causes a unitary transformation of the orbitals,
\begin{equation}
\hat{W}^{(1)} \hat{c}^\dagger_{\mathbf{i}_1}\hat{c}^\dagger_{\mathbf{i}_2} \cdots \hat{c}^\dagger_{\mathbf{i}_N} \ket{0} = \hat{c}^{\prime\prime\dagger}_{\mathbf{i}_1}\hat{c}^{\prime\prime\dagger}_{\mathbf{i}_2} \cdots \hat{c}^{\prime\prime\dagger}_{\mathbf{i}_N} \ket{0},\label{eqn:w1evolve}
\end{equation}
with $\hat{c}^{\prime\prime\dagger}_n = \sum_m c^{\dagger}_m W^{(1)}_{mn}$. This implies, by definition, that $\hat{W}^{(1)}$ does not affect the particle entanglement in the system. The orbital transformation can nevertheless affect the mode entanglement between two orbital partitions $A$ and $B$, unless $H^{(1)}$ and thus $\hat{W}^{(1)}$ is local in $A$ and $B$. Local orbital transformations do not affect the mode entanglement since $\hat{\rho}_A$ is invariant under any local unitary orbital transformation in $B$, and $S[\hat{\rho}_A]$ is independent of the unitary orbital transformations in $A$. A non-zero off-diagonal element in $W^{(1)}$ between $A$ and $B$ can easily affect the mode entanglement since it induces coherent transport of electrons between the two partitions. 

In the following we want to analyse how $\hat{W}^{(2)}$ affects the mode and particle entanglement. To this end, let us start with the derivation of Eq.~(\ref{eqn:w1evolve}) and then generalize it to $\hat{W}^{(2)}$.

The unitary operator $\hat{W}^{(1)}$ can be written as
\begin{equation}
\hat{W}^{(1)} = \exp[{i\sum_{mn} \frac{H^{(1)}_{mn}t}{M} \hat{c}^\dagger_{m} \hat{c}^{\phantom{\dagger}}_{n}}].
\end{equation}
The exponent can be brought to a diagonal form by diagonalizing $H^{(1)}$ using the eigenvectors $v_{mn}$ and the eigenvalues $E^{(1)}_n$,
\begin{equation}
\hat{W}^{(1)} = \exp({i\sum_{j} \frac{E^{(1)}_{j}t}{M} \hat{c}^{\prime\dagger}_j \hat{c}^{\prime}_j}),\label{eqn:diagonal1}
\end{equation}
where $\hat{c}^{\prime\dagger}_j = \sum_m c^{\dagger}_m v_{mj}$. Since the diagonal terms in the exponent commute, and $\hat{c}^{\prime\dagger}_j \hat{c}_j^{\prime} = \hat{c}^{\prime\dagger}_j \hat{c}_j^{\prime}\hat{c}^{\prime\dagger}_j \hat{c}_j^{\prime}$, $\hat{W}^{(1)}$ can be Taylor expanded as
\begin{equation}
\hat{W}^{(1)} = \prod_j \Big(1 - \hat{c}^{\prime\dagger}_j \hat{c}^{\prime}_j + e^{iE^{(1)}_j t/N} \hat{c}^{\prime\dagger}_j \hat{c}^{\prime}_j \Big).\label{eqn:taylor1}
\end{equation}
Given Eq.~(\ref{eqn:taylor1}) and that $\hat{c}^{\prime\dagger}_j \hat{c}^{\prime\dagger}_j = 0$, it follows that
\begin{equation}
\hat{W}^{(1)} \hat{c}^{\prime\dagger}_j = \hat{c}^{\prime\dagger}_j  e^{\frac{itE^{(1)}_j}{M}} \hat{W}^{(1)}.\label{eqn:w1cprime}
\end{equation}
Hence, if $\hat{W}^{(1)}$ acts on a creation operator $\hat{c}^{\dagger}_n$ from the left we get
\begin{align}
\hat{W}^{(1)} \hat{c}^{\dagger}_n &= \hat{W}^{(1)} \sum_j c^{\prime\dagger}_j v^{*}_{nj}\nonumber\\
&= \Big(\sum_j \hat{c}^{\prime\dagger}_j e^{iE^{(1)}_j t/M} v^{*}_{nj} \Big) \hat{W}^{(1)}\nonumber\\
&= \sum_{jkm} \hat{c}^{\prime\dagger}_k v^{*}_{mk} v_{mj} e^{iE^{(1)}_j t/M} v^{*}_{nj} \hat{W}^{(1)} \nonumber\\
&= \sum_{m}  \hat{c}^{\dagger}_{m} W^{(1)}_{mn} \hat{W}^{(1)},\label{eqn:w1steps}
\end{align}
where we in the third line used the Kronecker delta $\delta_{kj} = \sum_{m} v^{*}_{mk} v_{mj}$ and that
\begin{equation}
W^{(1)}_{mn} = (e^{iH^{(1)} t/M})_{mn} = \sum_{j} v_{mj} e^{iE^{(1)}_j t/M} v^{*}_{nj}.
\end{equation}
Eq.~(\ref{eqn:w1evolve}) follows immediately from Eq.~(\ref{eqn:w1steps}) and that $\hat{W}^{(1)} \ket{0} = \ket{0}$.

The Coulomb interaction $\hat{U}$ in Eq.~(\ref{eqn:u}) is diagonal in the position and spin basis $(r,\sigma,r',\sigma')$. We can therefore write $\hat{W}^{(2)}$ in a diagonal form, {c.f.} Eq.~(\ref{eqn:diagonal1}),
\begin{align}
\hat{W}^{(2)} & =  \exp[\frac{it}{M} \sum_{\sigma\sigma'}\iint \frac{t/N}{|r-r'|} \hat{c}^{\dagger}_{r\sigma} \hat{c}^{\dagger}_{r'\sigma'} \hat{c}^{\phantom{\dagger}}_{r'\sigma'} \hat{c}^{\phantom{\dagger}}_{r\sigma} dr'dr],\nonumber\\
& \equiv \exp[i\sum_{\mathbf{r} \in \mathcal{S}_2} \frac{E_{\mathbf{r}}t}{M} \hat{S}^{\dagger}_\mathbf{r} \hat{S}_\mathbf{r} ],
\end{align}
where the two-particle Slater determinant index $\mathbf{r}$ contains both position and spin, {i.e.} $\hat{S}_\mathbf{r} = \hat{c}_{r\sigma} \hat{c}_{r'\sigma'}$. The Taylor expansion of $\hat{W}^{(2)}$ becomes
\begin{equation}
\hat{W}^{(2)} = \prod_{\mathbf{r} \in \mathcal{S}_2} \Big(1 - \hat{S}_\mathbf{r}^{\dagger} \hat{S}_\mathbf{r} + e^{\frac{itE_\mathbf{r}}{M}} \hat{S}^{\dagger}_\mathbf{r}  \hat{S}_\mathbf{r} \Big),\label{eqn:w2taylor}
\end{equation}
which yields\footnote{It is straight forward to generalize Eq.~(\ref{eqn:w2cr}) to treat an effective $N$-body density-density interaction by replacing $r'$ and $\hat{c}_{r'}$ with an $N-1$ electron Slater determinant index $\mathbf{r}'$ and $\hat{S}_{\mathbf{r}'}$, respectivly.} 
\begin{align}
\hat{W}^{(2)} \hat{c}^{\dagger}_{r\sigma} & = \hat{c}^{\dagger}_{r\sigma} \!\!\!\!\!\!\!\!\!\! \prod_{(r\sigma r'\!\sigma') \in \mathcal{S}_{2}} \!\!\!\!\!\!\!\! [1 - \hat{c}^{\dagger}_{r'\!\sigma'} \hat{c}^{\phantom{\dagger}}_{r'\!\sigma'} + e^{\frac{itE_{(r\sigma r'\!\sigma')}}{M}} \hat{c}^{\dagger}_{r'\!\sigma'} \hat{c}^{\phantom{\dagger}}_{r'\!\sigma'}]  \hat{W}^{(2)} \nonumber \\
& = \hat{c}^{\dagger}_{r\sigma} \exp[{it}/{M}\!\!\!\!\!\!\!\sum_{(r\sigma r'\sigma') \in \mathcal{S}_{2}} \!\!\!\!\!\!\! E_{(r\sigma r'\sigma')} \hat{c}^{\dagger}_{r'\sigma'} \hat{c}^{\phantom{\dagger}}_{r'\sigma'}]  \hat{W}^{(2)} \nonumber\\
& \equiv  \hat{c}^{\dagger}_{r\sigma} \hat{W}^{(1)}_{r\sigma} \hat{W}^{(2)},\label{eqn:w2cr}
\end{align}
where $(r\sigma r'\sigma')$ denotes the Slater determinant index for which the electrons occupy $r$ and $r'$ with spin $\sigma$ and $\sigma'$, respectively. $\hat{W}^{(1)}_{r\sigma}$ is a one-particle unitary operator on the same form as $\hat{W}^{(1)}$ in Eq.~(\ref{eqn:diagonal1}) except that it depends on the position and spin of the creation operator $\hat{W}^{(2)}$ acted upon.

The main difference between Eq.~(\ref{eqn:w1cprime}) and Eq.~(\ref{eqn:w2cr}) is that the unitary operator $\hat{W}^{(1)}_{r\sigma}$ can not be absorbed by an orbital transformation. Instead, $\hat{W}^{(1)}_{r\sigma}$ will act upon the next creation operator $\hat{c}^\dagger_{r'\sigma'}$ in line giving
\begin{equation}
\hat{W}^{(2)}_{r\sigma} \hat{c}^{\dagger}_{r'\sigma'}  = \hat{c}^{\dagger}_{r'\sigma'} e^{i E_{(r\sigma r'\sigma')} t/M}  \hat{W}^{(2)}_{r\sigma}.\label{eqn:w2crprime}
\end{equation}
Eq.~(\ref{eqn:w2cr}) and (\ref{eqn:w2crprime}) fully determine the evolution given by $\hat{W}^{(2)}$,
\begin{align}
\hat{W}^{(2)} \hat{c}^\dagger_{\mathbf{r}_1}\hat{c}^\dagger_{\mathbf{r}_2} \cdots \hat{c}^\dagger_{\mathbf{r}_N} \ket{0}
& = \hat{c}^\dagger_{\mathbf{r}_1}\hat{W}^{(2)}_{\mathbf{r}_1} \hat{c}^\dagger_{\mathbf{r}_2} \hat{W}^{(2)}_{\mathbf{r}_1} \cdots \hat{c}^\dagger_{\mathbf{r}_N} \ket{0}\nonumber\\
& = \prod_{j=1}^N \hat{c}^\dagger_{\mathbf{r}_j} e^{\frac{it}{2M} \sum_{n \neq j}^N E_{(\mathbf{r}_n\mathbf{r}_i)}} \ket{0}.\label{eqn:wevolve}
\end{align}
The phase factors induced by $\hat{W}^{(2)}$ in Eq.~(\ref{eqn:wevolve}) depend non-linearly on the orbital occupations. 
It is therefore not possible to assign to each orbital a fixed phase shift as in Eq.~(\ref{eqn:w1cprime}), unless the many-body state is an eigenstate to $\hat{U}$.
Hence, in contrast to one-particle terms in the Hamiltonian, the evolution given by the Coulomb interaction will in general modify the particle entanglement. 
The mode entanglement between two orbital partitions $A$ and $B$ is also in general affected by $\hat{W}^{(2)}$. However, just as in the one-particle case, a change in the mode entanglement requires off-diagonal phases between $A$ and $B$, {i.e.} that the electrons in $A$ and $B$ interact with each other. 

These effects can be illustrated with a minimal model; a system with two orbitals $1$ and $2$ and two electrons, with $H^{(1)} = 0$ and an on-site effective Coulomb interaction of the form 
\begin{equation}
\hat{U} = U \hat{c}^\dagger_{1\uparrow}\hat{c}^\dagger_{1\downarrow}\hat{c}_{1,\downarrow}\hat{c}_{1\uparrow} + U\hat{c}^\dagger_{2\uparrow}\hat{c}^\dagger_{2\downarrow}\hat{c}_{2\downarrow}\hat{c}_{2\uparrow}. \label{eqn:uexample}
\end{equation}
The two electrons are prepared in the Fock state $\ket{\Psi} = \hat{c}^\dagger_{a\uparrow}\hat{c}^\dagger_{b\downarrow} \ket{0}$ where
\begin{equation}
\hat{c}_{a\sigma}^{\dagger} = \frac{1}{\sqrt{2}}(\hat{c}_{1\sigma}^{\dagger}+\hat{c}_{2\sigma}^{\dagger}) 
\hspace{2em}
\hat{c}_{b\sigma}^{\dagger} = \frac{1}{\sqrt{2}}(\hat{c}_{1\sigma}^{\dagger}-\hat{c}_{2\sigma}^{\dagger}).\label{eqn:bonding}
\end{equation}
The state $\ket{\Psi}$ evolve according to Eq.~(\ref{eqn:wevolve}), 
\begin{align}
W^{(2)}\ket{\Psi} & = \frac{W^{(2)}(t)}{2}\Big(\hat{c}^\dagger_{1\uparrow}\hat{c}^\dagger_{1\downarrow}\!  -\hat{c}^\dagger_{1\uparrow} \hat{c}^\dagger_{2\downarrow}\! + \hat{c}^\dagger_{2\uparrow}\hat{c}^\dagger_{1\downarrow}\! - \hat{c}^\dagger_{2\uparrow}\hat{c}^\dagger_{2\downarrow} \Big)\ket{0}\nonumber\\
 = &~ \frac{1}{2}\Big(e^{i U t}\hat{c}^\dagger_{1\uparrow}\hat{c}^\dagger_{1\downarrow}\!  -\hat{c}^\dagger_{1\uparrow} \hat{c}^\dagger_{2\downarrow}\! + \hat{c}^\dagger_{2\uparrow}\hat{c}^\dagger_{1\downarrow}\! - e^{i U t}\hat{c}^\dagger_{2\uparrow}\hat{c}^\dagger_{2\downarrow}\Big) \ket{0} \nonumber\\
 = &\, \Big(\frac{e^{i U t} + 1}{2} \hat{c}^\dagger_{a\uparrow}\hat{c}^\dagger_{b\downarrow} + \frac{e^{i U t} - 1}{2} \hat{c}^\dagger_{a\downarrow}\hat{c}^\dagger_{b\uparrow}\Big)\ket{0}.
\end{align}
As $\ket{\Psi}$ evolves it goes from a Fock state at $t=0$ to become maximally particle entangled at $t=\frac{\pi}{2U}$. The mode entanglement between orbitals $a$ and $b$ changes at the same time from $S[\rho_a(t=0)] = 0$ to $S[\rho_a(t=\frac{\pi}{2U})] = 1/2$. The mode entanglement between $1$ and $2$ is however not affected by the on-site effective Coulomb interaction since it is local with respect to $1$ and $2$.

\subsection{Particle entanglement and measurements}
The particle entanglement is not only affected by the interactions within a system but also by the interaction with a measurement device. The underlying principle of any measurement device is a non-linear amplification process that is triggered by its interaction with the probed system. The amplification process is in general a chain reaction designed to correlate the states of a macroscopically large number of particles with the state of the probed system. 
For example, an electron multiplier measures the occupation of a free electron orbital ($e$) at time $t=0$ by accelerating the (primary) electron towards a surface of a secondary-emissive material. The interaction with the surface electrons (${D}^0$) cause the ejection of secondary electrons which in turn are accelerated toward a second surface where the process is repeated. The chain reaction cause a cascade of excited electrons (${D}^1$) to travel down the electron multiplier to finally, at time $t=T$, reach an electric readout. An electron multiplier can in general detect the presence of an electron in more than a single spin-orbital. This can however be treated as an array of detectors each detecting just a single spin-orbital but having a common electrical readout. The ideal one-particle detection process can hence be summarized as 
\begin{align}
\ket{\Psi^i} = & \hat{c}^\dagger_{e} \Big(\sum_{\mathbf{i} \in  \mathcal{S}_{N}} D^0_{\mathbf{i}}(0) \hat{S}^\dagger_{\mathbf{i}} \Big)\ket{0} \nonumber \\
\rightarrow \ket{\Psi^f} = & \Big(\sum_{\mathbf{i} \in  \mathcal{S}_{N+1}} D^1_{\mathbf{i}}(T) \hat{S}^\dagger_{\mathbf{i}} \Big) \ket{0}.\label{eqn:detector}
\end{align}
The pointer states of the detector (${D}^0$ and ${D}^1$) are at equal time $t = T$ for all practical purposes irreversibly orthogonal to each other, {i.e.}
\begin{equation}
\bra{0} \Big(\sum_{\mathbf{i}} D^{0*}_{\mathbf{i}}(T) \hat{S}_{\mathbf{i}} \Big) \hat{c}^\dagger_{s} \hat{W}(t') \Big(\sum_{\mathbf{i}} D^1_{\mathbf{i}}(T) \hat{S}^\dagger_{\mathbf{i}} \Big) \ket{0} = 0
\end{equation}
where $\hat{W}(t')$ is any practically achievable evolution operator, and the time $t'$ can be made arbitrarily large by increasing the size of the detector. To highlight this irreversibility we will in the following use a notation that separates the state of the probed system ($\Psi$) and the detector ($D$), 
\begin{equation}
\ket{\Psi,D^0(0)} \equiv \Big(\sum_{\mathbf{i}} \Psi_{\mathbf{i}} \hat{S}^\dagger_{\mathbf{i}}\Big) \Big(\sum_{\mathbf{i}} D^0_{\mathbf{i}}(0) \hat{S}^\dagger_{\mathbf{i}} \Big) \ket{0}. 
\end{equation}

The interaction between the system and the detector in Eq.~(\ref{eqn:detector}) causes in general the formation of a superposition between the different pointer states, 
\begin{align} 
\ket{\Psi^i} \equiv & \sqrt{\alpha}e^{i\theta}\hat{c}^\dagger_{e}\ket{\Psi_e,D^0(0)} + \sqrt{1-\alpha}\ket{\Psi_s,D^0(0)} \rightarrow \nonumber \\
\ket{\Psi_f} \equiv & \sqrt{\alpha}e^{i\theta}\ket{\Psi_e,D^1(T)} + \sqrt{1-\alpha}\ket{\Psi_s,D^0(T)},\label{eqn:entdetector}
\end{align}
where $0 \le \alpha \le 1$ and the state $\ket{\Psi_s}$ does not trigger the detector. However, since the different pointer states of the detector always remain orthogonal to each other it is impossible to access the relative phase $e^{i\theta}$ in $\ket{\Psi^f}$. The irreversible lack of interference makes the superposition between the different pointer states equivalent to a statistical (classical) correlation. The particle entanglement within the detector, {i.e.} $E[\ket{0,D^0}]$ and $E[\ket{0,D^1}]$, is in general equally unaccessible. In the following we will therefore focus on the particle entanglement within the probed system ($S$) given by $E[\hat{P}_{D^p} \ket{\Psi^f}]$, where 
the projection operator
\begin{equation}
\hat{P}_{D^p} \equiv \sum_{\substack{\mathbf{i} \in \mathcal{S}^D\\ \mathbf{j} \in \mathcal{S}^S}} D^{p*}_{\mathbf{i}}\hat{P}^{S}_{\mathbf{j},\mathbf{i}}
\end{equation}
resolves and projects out the pointer state $D^p$ of the detector using the projectors $\hat{P}_{S,\mathbf{i}}$ defined by Eq.~(\ref{eqn:projection}). 

As shown in Appendix \ref{app:entmeasure}
the particle entanglement of the probed system in Eq.~(\ref{eqn:entdetector}) fulfills the inequality
\begin{equation}
E[\hat{P}_{D^0} \ket{\Psi^i}] \ge E[\hat{P}_{D^0} \ket{\Psi^f}] + E[\hat{P}_{D^1} \ket{\Psi^f}],\label{eqn:entrelation}
\end{equation}
for both $E_G[\ket{\Psi}]$ and $S[\ket{\Psi}]$. 
Eq.~(\ref{eqn:entrelation}) can be rewritten as
\begin{align}
E\Big[\sqrt{\alpha}e^{i\theta}\hat{c}^\dagger_{e}\ket{\Psi_e} &+ \sqrt{1-\alpha}\ket{\Psi_s}\Big] \nonumber\\
& \ge \alpha E\Big[\ket{\Psi_e}\Big] + (1-\alpha)E\Big[\ket{\Psi_s}\Big],\label{eqn:entrelation2}
\end{align}
which shows that the measurement of the occupation of a single spin-orbital does not increase the particle entanglement of the probed system on average. 
This parallels how local measurements and classical communiction do not increase the entanglement of distinguishable particles on average \cite{Vedral_PRL_1997_ent_meas}. 

The interaction with the detector is often preceded by an excitation of the probed system. In photoemission spectroscopy the system is first excited by a photon which causes an electron to be emitted from the surface of the sample. The electron enters a highly excited free electron orbital and is then captured by an electron multiplier as in Eq.~(\ref{eqn:entdetector}). If the excitation by the photon is considered to be much faster than the dynamics of the system then it can be described by an effective one-particle unitary operator ($\hat{W}^{(1)}$). This implies that the full ideal photoemission spectroscopy measurement also fulfills Eq.~(\ref{eqn:entrelation}) and not only the detection step. 

The working principle of a photomultiplier is the same as an electron multiplier except that the primary particle is a photon instead of an electron. The incoming photon hits a photo-emissive surface which in the ideal case cause the emission of a free electron. The free electron is then detected by an electron multiplier according to Eq.~(\ref{eqn:entdetector}). Although the detection of the free electron fulfills the particle entanglement inequality in Eq.~(\ref{eqn:entrelation}), the detection of a photon may still cause the formation of particle entanglement. The reason is that the combined emission and absorbtion of the photon corresponds to an interaction which may cause particle entanglement to form within the probed system according to Eq.~(\ref{eqn:wevolve}). 

The projective quantum non-demolition measurement\cite{braginsky1980quantum,nogues1999seeing} in the definition of the N-SSR restricted mode entanglement 
is only probing the total number of electrons in an orbital partition and not the occupation of individual spin-orbitals. This can in general only be achieved by letting the electrons in the probed system interact with an auxiliary system, and then measure the state of the auxiliary system. The interaction can increase the average particle entanglement of the probed system, which is the physical reason why a mode entangled Fock state can become particle entangled after the projective measurement.

\section{Fermionic teleportation}
\label{sec:examples}
\subsection{Electron teleportation in a hydrogen molecule} 
\label{Sec:1ex}

\begin{figure*}[t!]
\begin{minipage}{18. cm}
	\begin{subfigure}{8.5 cm}
  		\centering
		\includegraphics[width=7.cm]{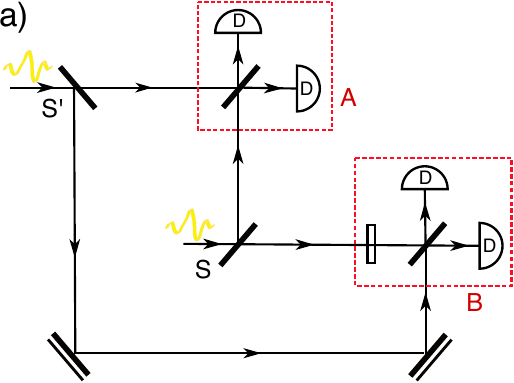}
	\end{subfigure}
	\begin{subfigure}{8.5 cm}
  		\centering
  		\includegraphics[clip=true,trim=50 220 200 0,width=7.cm]{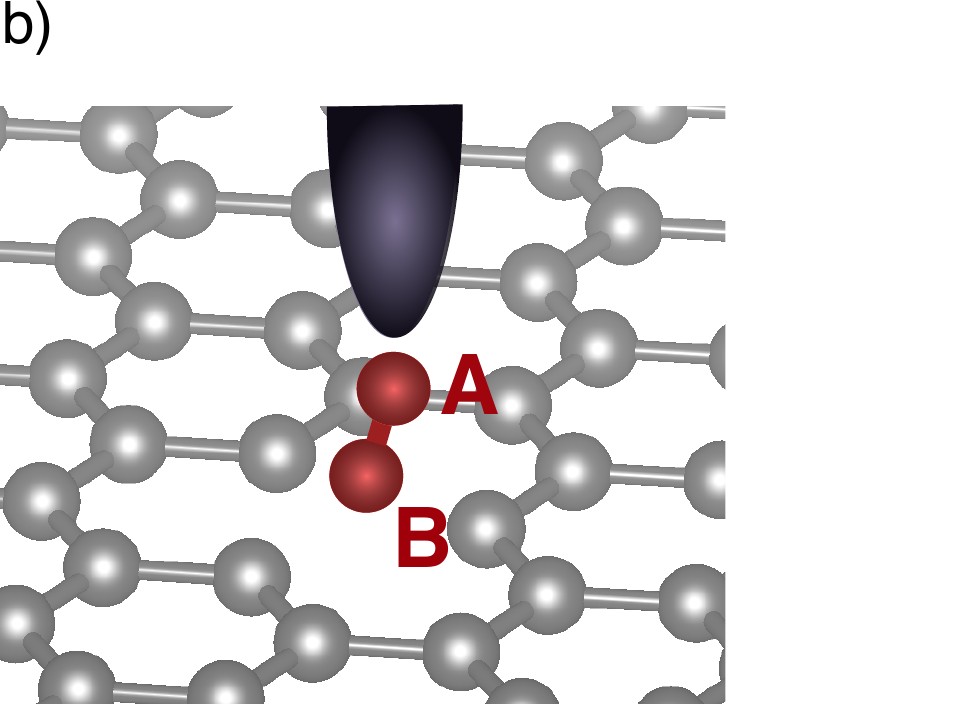}
	\end{subfigure}
	\caption{a) Photonic and b) electronic teleportation schemes. a) In the photon teleportation scheme of Ref.\onlinecite{Lee_2photon_tele} an entangled resource state shared between \textit{A(lice)} and \textit{B(ob)} is created through a single photon entering a 50/50 beam-splitter $S$. The state to be teleported instead is generated at $S'$.  \textit{A(lice)} performs a projective measurement on her part of the resource state and the state to be teleported ($D$ are detectors). Depending on the outcome of this measurement, \textit{B(ob)} needs to perform a suitable operation (phase shift) on his state in order to obtain the teleported state. b) In our electron teleportation scheme the two H-atoms of an H$_2$ molecule play the role of \textit{A(lice)} and \textit{B(ob)}. The H$_2$ molecule is adsorbed on a single vacancy in graphene and can be addressed by a spin-polarized STM-tip. The latter can be used to add an electron to the molecule (state to be teleported, see Eq.(\ref{eq:2e_tele})) and to perform a magnetization measurement.}
	\label{fig:2e_teleportation}
\end{minipage}
\end{figure*}
Our first teleportation scheme can be seen as a molecular analog of the photon teleportation scheme presented in Ref.~\onlinecite{Lee_2photon_tele} which is schematically depicted in Fig.~\ref{fig:2e_teleportation}a. In this example we consider a hydrogen molecule that is adsorbed on graphene and stabilized by a magnetic scanning tunneling microscope (STM) tip, as shown in Fig.~\ref{fig:2e_teleportation}b. The hydrogen molecule is adsorbed on a single vacancy site since this adsorption site is favoured by its binding energy ($\approx 0.4eV$) \cite{graphene_defect_eng}. In Fig.~\ref{fig:2e_teleportation}b the two constituting hydrogen atoms H of the molecule are labelled $A$ and $B$, respectively, and will play the role of $Alice$ and $Bob$. We assume that the molecule is initially in its ionized state H$_2^+$, with one spin $\downarrow$ electron in its binding orbital $\sigma$ (with the spin pointing in the in-plane $x$-direction). Thus, the initial wave function of the molecule reads           
\begin{equation}
\label{eq:init_state}
\hat{c}^\dagger_{\sigma\downarrow}\ket{0} = \frac{1}{\sqrt{2}}(\hat{c}^\dagger_{A\downarrow}+\hat{c}^\dagger_{B\downarrow})\ket{0},
\end{equation}
where $\hat{c}^\dagger_{\sigma\downarrow}$ ($\hat{c}^\dagger_{A\downarrow}$ and $\hat{c}^\dagger_{B\downarrow}$)  create an electron with spin $\downarrow$  in the binding orbital $\sigma$ which is a superposition of the $s$-orbitals of H-atoms $A$ and $B$. 
$\hat{c}^\dagger_{\sigma\downarrow}\ket{0}$  represents the initial resource state shared between $A(lice)$ and $B(ob)$.  Its mode entanglement, given the bipartition between $A$ and $B$, can be calculated from the reduced density matrix $\rho_A$, defined in Eq.~\eqref{eq:red_dmatrix}. $\rho_A$, written in the basis $\{\ket{0},\hat{c}^\dagger_{A\downarrow}\ket{0},\hat{c}^\dagger_{A\uparrow}\ket{0},\hat{c}^\dagger_{A\downarrow}\hat{c}^\dagger_{A\uparrow}\ket{0}\}$, reads 
\begin{equation}
\rho_A=\frac{1}{2}
\begin{pmatrix}
1&0&0&0\\
0&1&0&0\\
0&0&0&0\\
0&0&0&0\\
\end{pmatrix}
\end{equation}
which gives $S[\rho_A] = \Tr[\rho_A(1-\rho_A)]=1/2$. Clearly, $\hat{c}^\dagger_{\sigma\downarrow}\ket{0}$ is not particle entangled as it only contains a single electron.

Next, we inject a second electron with spin $\uparrow$ into the H$_2^+$ molecule via the spin-polarized STM tip. The energy of this injected electron is chosen so that it occupies a superposition $\gamma$ of the binding $\sigma$ and anti-binding $\bar{\sigma}$ orbital of the H$_2$ molecule, namely   
\begin{equation}
\label{eq:2e_tele}
\hat{c}^\dagger_{\gamma\uparrow}\ket{0} = [\frac{a+b}{\sqrt{2}}\hat{c}^\dagger_{\sigma\uparrow}+\frac{a-b}{\sqrt{2}}\hat{c}^\dagger_{\bar{\sigma}\uparrow}]\ket{0} =
(a\hat{c}^\dagger_{A\uparrow}+b\hat{c}^\dagger_{B\uparrow})\ket{0},
\end{equation}
where $a$ and $b$ are unknown coefficients, and $|a|^2+|b|^2=1$. $\hat{c}^\dagger_{\gamma\uparrow}\ket{0}$ corresponds to the unknown state to be teleported. Its mode entanglement can be calculated from its reduced density matrix $\rho_A$
\begin{equation}
\rho_A=
\begin{pmatrix}
|b|^2&0&0&0\\
0&0&0&0\\
0&0&|a|^2&0\\
0&0&0&0\\
\end{pmatrix}
\end{equation}
and yields $S[\rho_A]=1-|a|^4-|b|^4$. $\hat{c}^\dagger_{\gamma\uparrow}\ket{0}$ is again not particle entangled since it only contains a single electron.

The total state of the H$_2$ molecule is given by       
\begin{equation}\label{state_2e}
\ket{\psi} \equiv \hat{c}^\dagger_{\sigma\downarrow}\hat{c}^\dagger_{\gamma\uparrow}\ket{0} = \frac{1}{\sqrt{2}}(\hat{c}^\dagger_{A\downarrow}+\hat{c}^\dagger_{B\downarrow})(a\hat{c}^\dagger_{A\uparrow}+b\hat{c}^\dagger_{B\uparrow})\ket{0}.
\end{equation}
$\ket{\psi}$ is mode entangled if we consider a bipartition between hydrogen atoms $A$ and $B$. In this case, the reduced density matrix $\rho_A$ becomes 
\begin{equation}
\rho_A= \frac{1}{2}
\begin{pmatrix}
|b|^2&0&0&0\\
0&|a|^2&0&0\\
0&0&|b|^2&0\\
0&0&0&|a|^2\\
\end{pmatrix}
\end{equation}
which yields the entropic mode entanglement $S[\rho_A]=1-1/2(|a|^4+|b|^4)$. Note that this is not simply the sum of the entropic entanglement of the states in Eqs.~\eqref{eq:init_state} and \eqref{eq:2e_tele} since we are considering linear and not logarithmic entropy. $\ket{\psi} = \hat{c}^\dagger_{\sigma\downarrow}\hat{c}^\dagger_{\gamma\uparrow}\ket{0}$ is not a particle-entangled state as it clearly can be written as a single Slater determinant. 
 
The next step in our teleportation scheme is a measurement of the spin of the electrons in atom $A$ along the $y$-axis. Even if very challenging, this measurement can in principle be performed by exciting the electrons in $A$ with a photon to make them tunnel to the magnetized STM-tip which detects their spin. This causes the build up of particle entanglement between the system and the detector, but it does not increase the particle entanglement within the probed system itself as the spin measurement corresponds to two separate spin-orbital occupation measurements which fulfill Eq.~(\ref{eqn:entrelation}).

To analyze the possible outcomes of the measurement, we substitute $\hat{c}^\dagger_{A\uparrow(\downarrow)}= 1/\sqrt{2}\left(\hat{c}^\dagger_{A\uparrow y}\varpm\hat{c}^\dagger_{A\downarrow y}\right)$ and rewrite Eq.~(\ref{state_2e}) as
\begin{align}
\label{state_2e_expl}
\ket{\psi}  & = \frac{1}{\sqrt{2}}\left[\frac{1}{\sqrt{2}}(\hat{c}^\dagger_{A\uparrow y}-\hat{c}^\dagger_{A\downarrow y})+\hat{c}^\dagger_{B\downarrow}\right]\times  \nonumber\\& \times \left[\frac{a}{\sqrt{2}}(\hat{c}^\dagger_{A\uparrow y}+\hat{c}^\dagger_{A\downarrow y})+b\hat{c}^\dagger_{B\uparrow}\right]\ket{0} \nonumber\\&
= \frac{1}{\sqrt{2}}[\frac{\hat{c}^\dagger_{A\uparrow y}}{\sqrt{2}}(-a\hat{c}^\dagger_{B\downarrow}+b\hat{c}^\dagger_{B\uparrow}) -  \frac{\hat{c}^\dagger_{A\downarrow y}}{\sqrt{2}}(a\hat{c}^\dagger_{B\downarrow}+b\hat{c}^\dagger_{B\uparrow})+ \nonumber\\& + a\hat{c}^\dagger_{A\uparrow y}\hat{c}^\dagger_{A\downarrow y}+ b\hat{c}^\dagger_{B\downarrow}\hat{c}^\dagger_{B\uparrow} ]\ket{0} .    
\end{align}

If Alice either detects no electrons (due to $\frac{b}{\sqrt{2}}\hat{c}^\dagger_{B\downarrow}\hat{c}^\dagger_{B\uparrow}\ket{0}$) or two electrons (due to $\frac{a}{\sqrt{2}}\hat{c}^\dagger_{A\uparrow y}\hat{c}^\dagger_{A\downarrow y}\ket{0}$) from atom $A$ then no teleportation can be achieved, as Bob's spin-orbitals are then either completely filled or empty, respectively, due to particle number conservation. Thus, probability of a successful teleportation is limited to 50\%. % This is a total occupation measurement...

This limited success rate is reflected in the particle number superselection (N-SSR) restricted mode entanglement of the state in Eq.~\eqref{state_2e_expl}. Indeed, the only part of $\ket{\psi}$  which contributes to the N-SSR restricted mode entanglement   are the terms with one electron at A and B, respectively, i.e.   
\begin{equation}
\label{final_2e}
\ket{\psi^{(1)}} = \left[\frac{\hat{c}^\dagger_{A\uparrow y}}{\sqrt{2}}(a\hat{c}^\dagger_{B\downarrow}-b\hat{c}^\dagger_{B\uparrow}) +  \frac{\hat{c}^\dagger_{A\downarrow y}}{\sqrt{2}}(a\hat{c}^\dagger_{B\downarrow}+b\hat{c}^\dagger_{B\uparrow})\right]\ket{0} 
\end{equation}
The reduced density matrix of $\ket{\psi^{(1)}}$ reads
\begin{equation}
\rho_A= 
\begin{pmatrix}
0&0&0&0\\
0&|b|^2&0&0\\
0&0&|a|^2&0\\
0&0&0&0\\
\end{pmatrix}
\end{equation}
and yields a standard mode entanglement of $S[\rho_A]=1-|a|^4-|b|^4$. For the N-SSR restricted mode entanglement of $\ket{\psi}$ this value is further rescaled by the probability of measuring 1 electron in A and B, respectively. This finally yields for the N-SSR restricted mode entanglement of $\ket{\psi}$
\begin{equation}
S^{N}(\ket{\psi}) = \frac{1}{2}(1-|a|^4-|b|^4).
\end{equation}
  
The two successful magnetization measurement outcomes of $\ket{\psi}$  are $\frac{1}{2}\hat{c}^\dagger_{A\uparrow y}(-a\hat{c}^\dagger_{B\downarrow}+b\hat{c}^\dagger_{B\uparrow})\ket{0}$ and $\frac{1}{2}\hat{c}^\dagger_{A\downarrow y}(a\hat{c}^\dagger_{B\downarrow}+b\hat{c}^\dagger_{B\uparrow})\ket{0}$. In the latter case the electron on atom $B$ is already in the state to be teleported (see Eq.~\eqref{eq:2e_tele}), and no further operation is necessary to complete the teleportation scheme. In the former case Bob still needs to apply a 'phase shift' to the electron spin on atom $B$ in order to obtain the desired outcome.
This can be achieved by applying a magnetic field on atom $B$ in $\downarrow x$-direction for a specific time $t=\pi/(\mu B_x)$, which yields
\begin{align}
e^{-\frac{i}{\hbar}\mu B_x\hat{S}_xt}(a\hat{c}^\dagger_{B\downarrow}-b\hat{c}^\dagger_{B\uparrow})\ket{0} &=( e^{-i\frac{\pi}{2}}a\hat{c}^\dagger_{B\downarrow}-e^{i\frac{\pi}{2}}b\hat{c}^\dagger_{B\uparrow}) \ket{0} \nonumber \\ 
& = -i(a\hat{c}^\dagger_{B\downarrow}+b\hat{c}^\dagger_{B\uparrow})\ket{0}.
\end{align}

This example shows that it is possible to teleport a state without creating or destroying particle entanglement {\em within} the system. The caveat is that there is only a 50\% probability of success. This fact is reflected by a reduction of the mode entanglement through the particle-number superselection rule (N-SSR). 
A 100\% success rate requires the use of Bell states\cite{reliable_tele_1998}, which are maximally particle and mode entangled by construction, as illustrated in the next example.

\subsection{NV-center teleportation scheme}
\label{Sec:2ex}
\begin{figure}[]
\includegraphics[width=5.cm]{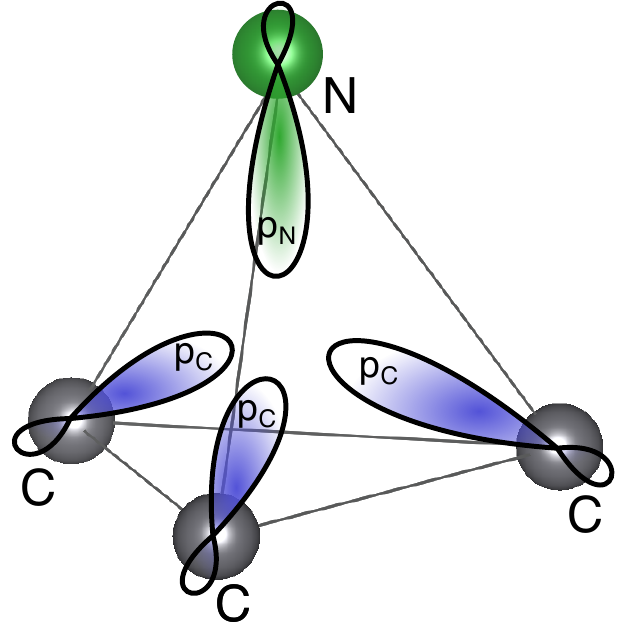}
\caption{Schematic visualization of a NV center in diamond and its dangling $p$-orbitals oriented towards the vacancy (in the center of the pyramid). The N-$p$ orbital is labelled with $p_N$, while the corresponding C-$p$ orbitals are labelled with $p_C$.}
\label{fig:NV_sketch}
\end{figure}

In our second example we consider a nitrogen-vacancy (NV) center in diamond --- a point defect in the diamond lattice which consists of a substitutional N atom and a neighboring vacancy site (see Fig.~\ref{fig:NV_sketch}). NV centers are promising candidates for implementing quantum technologies since they exhibit atom-like properties in a solid-state environment.\cite{childress_hanson_2013,Weber8513}  It is their long-lived spin quantum states which can be addressed via optical transitions that make them particularly attractive as solid-state spin qubits. Numerous experiments involving NV centers have successfully been carried out in recent years,\cite{Wrachtrup2001,Astner2018_nv,liu_nature2019_nv} among them a quantum teleportation between distant NV centers.\cite{Pfaff_science_nv_tele} Here, we investigate a teleportation scheme within a single neutral NV$^0$ center. While experiments usually concentrate on the negatively charged NV$^-$, also the neutral NV$^0$ has recently attracted attention.\cite{Gali_PRB2009_nv0,Ranjbar_PRB2011_nv0,Peng_PRB2018_nv0} 
The teleportation scheme is inspired by the work in Ref.~\onlinecite{reliable_tele_1998}, where methods for reliable teleportation involving interactions among the involved quantum particles are investigated. 

Our proposed NV$^0$ teleportation scheme involves five electrons in eight spin-orbitals, which corresponds to three holes residing in the dangling-bond orbitals around a vacancy site in diamond. More specifically three dangling $p$-orbitals with angular momentum $m_l=-1,0,1$ localized on the three C-atoms nearest to the vacancy site and one dangling $p$-orbital with angular momentum $m_l=0$ localized on the N-atom (see Fig.~\ref{fig:NV_sketch} for a schematic visualization). We start out with an entangled pair of holes in  NV$^0$. We assume that this pair of holes is prepared in the following Bell state
\begin{equation}
\label{eq:NV0_init}
\ket{\psi_1} = \frac{1}{\sqrt{2}}(\hat{c}_{N\downarrow}\hat{c}_{1\downarrow} + \hat{c}_{N\uparrow}\hat{c}_{-1\downarrow})\ket{\bold{1}},
\end{equation}
where $\hat{c}_i$ are annihilation operators which annihilate an electron, i.e. create a hole, in the respective spin-orbital $i$. For simplicity, the annihilation operators for the C spin-orbitals do not carry a subscript $C$. $\ket{\bold{1}}$ represents the filled state with fully occupied spin-orbitals. $\ket{\psi_1}$ is a particle entangled state, its maximum overlap with a Fock state is 1/2 which yields a geometric entanglement measure of $E_G=1-1/2=1/2$. Furthermore, from its one-particle reduced density matrix, which in the basis \{N$\downarrow$,N$\uparrow$,1$\downarrow$,-1$\downarrow$\} reads
\begin{equation}\label{eq:dmatrix_1}
\rho^{(1\mathrm{p})}= \frac{1}{2}
\begin{pmatrix}
1&0&0&0\\
0&1&0&0\\
0&0&1&0\\
0&0&0&1\\
\end{pmatrix},
\end{equation}
we can calculate an entropic particle entanglement of $S[\rho^{(1\mathrm{p})}] =1$. On the other hand, if we introduce a bipartition between the C orbitals (\textit{Alice}) and the N orbitals (\textit{Bob}), we obtain a reduced density matrix (in the basis $\{\ket{0},\hat{c}^\dagger_{N\downarrow}\ket{0},\hat{c}^\dagger_{N\uparrow}\ket{0},\hat{c}^\dagger_{N\downarrow}\hat{c}^\dagger_{N\uparrow}\ket{0}\}$) 
\begin{equation}
\rho_N=\frac{1}{2}
\begin{pmatrix}
0&0&0&0\\
0&1&0&0\\
0&0&1&0\\
0&0&0&0\\
\end{pmatrix},
\end{equation}
which yields an entropic mode entanglement of $S[\rho_N]=1/2$. 
Thus --- unlike in our first example --- here we already start out with a maximally particle and mode entangled state $\ket{\psi_1}$. 

The state to be teleported is initially encoded in the spin state of a third hole residing in the C,$m_l=0$ orbital. The state of this hole reads     
\begin{equation}
\label{eq:NV0_tele}
\ket{\psi_2} = (a\hat{c}_{0\uparrow}+b\hat{c}_{0\downarrow})\ket{\bold{1}},
\end{equation}
where $a$ and $b$ are unknown coefficients, and $|a|^2+|b|^2=1$.  Clearly, $\ket{\psi_2}$ is neither particle entangled nor mode entangled with respect to the bipartition between the N and the C orbitals.
By combining $\ket{\psi_1}$ and $\ket{\psi_2}$, we obtain the total state of the NV$^0$ which reads 
\begin{equation}
\label{eq:NV_state_tot}
\ket{\psi}  = \frac{1}{\sqrt{2}}(\hat{c}_{N\downarrow}\hat{c}_{1\downarrow} + \hat{c}_{N\uparrow}\hat{c}_{-1\downarrow}) (a\hat{c}_{0\uparrow}+b\hat{c}_{0\downarrow})\ket{\bold{1}} .  
\end{equation}
The particle and mode entanglement of this state are the same as for $\ket{\psi_1}$ in Eq.~\eqref{eq:NV0_init} since $\ket{\psi_2}$ does not add any entanglement. 
We can rewrite Eq.~\eqref{eq:NV_state_tot} in the following form
\begin{align}
\label{eq:NV_state_tot_1}
\ket{\psi} & = \frac{1}{2\sqrt{2}}(\hat{c}_{1\downarrow}\hat{c}_{0\uparrow} - \hat{c}_{-1\downarrow}\hat{c}_{0\downarrow})(a\hat{c}_{N\downarrow}-b\hat{c}_{N\uparrow})\ket{\bold{1}} \nonumber\\ 
& + \frac{1}{2\sqrt{2}} (\hat{c}_{1\downarrow}\hat{c}_{0\uparrow} + \hat{c}_{-1\downarrow}\hat{c}_{0\downarrow})(a\hat{c}_{N\downarrow}+b\hat{c}_{N\uparrow})\ket{\bold{1}} \nonumber\\
& + \frac{1}{2\sqrt{2}} (\hat{c}_{-1\downarrow}\hat{c}_{0\uparrow} - \hat{c}_{1\downarrow}\hat{c}_{0\downarrow})(a\hat{c}_{N\uparrow}-b\hat{c}_{N\downarrow})\ket{\bold{1}} \nonumber\\
& + \frac{1}{2\sqrt{2}}(\hat{c}_{-1\downarrow}\hat{c}_{0\uparrow} + \hat{c}_{1\downarrow}\hat{c}_{0\downarrow})(a\hat{c}_{N\uparrow}+b\hat{c}_{N\downarrow})\ket{\bold{1}} ,
\end{align}
where we have factorized every contribution to $\ket{\psi}$ into two parts (brackets). The second brackets clearly resemble, apart from phase factors (signs), the original state to be teleported of Eq.~\eqref{eq:NV0_tele}. However, in oder to perform the teleportation, we need to be able to distinguish between the four lines of Eq.~\eqref{eq:NV_state_tot_1} through a measurement. For this purpose we consider a Coulomb interaction among the holes residing in the  C  spin-orbitals. We focus on spin-flip processes induced by the Coulomb interaction and disregard other processes. Let us first recall that the Coulomb interaction conserves the total spin and angular momentum. Thus, terms with same spins, such as $\hat{c}_{-1\downarrow}\hat{c}_{0\downarrow}$, remain unchanged. Terms with opposite spins instead  can undergo the following spin-flip transitions   
\begin{align}
\label{eq:spin_flip}
\hat{c}_{1\downarrow}\hat{c}_{0\uparrow} & \longrightarrow \hat{c}_{1\uparrow}\hat{c}_{0\downarrow} \nonumber \\
\hat{c}_{-1\downarrow}\hat{c}_{0\uparrow} & \longrightarrow \hat{c}_{-1\uparrow}\hat{c}_{0\downarrow},
\end{align}
where all spins of the involved spin-orbitals are flipped.  By applying these spin-flip Coulomb interactions to $\ket{\psi}$, we obtain
\begin{align}
\label{eq:NV_state_tot_after}
\ket{\psi'} & = -\frac{1}{2\sqrt{2}}\hat{c}_{0\downarrow}(\hat{c}_{1\uparrow} - \hat{c}_{-1\downarrow})(a\hat{c}_{N\downarrow}-b\hat{c}_{N\uparrow})\ket{\bold{1}} \nonumber\\ 
& -\frac{1}{2\sqrt{2}}\hat{c}_{0\downarrow}(\hat{c}_{1\uparrow} + \hat{c}_{-1\downarrow})(a\hat{c}_{N\downarrow}+b\hat{c}_{N\uparrow})\ket{\bold{1}} \nonumber\\
& - \frac{1}{2\sqrt{2}}\hat{c}_{0\downarrow} (\hat{c}_{-1\uparrow} - \hat{c}_{1\downarrow})(a\hat{c}_{N\uparrow}-b\hat{c}_{N\downarrow})\ket{\bold{1}} \nonumber\\
& - \frac{1}{2\sqrt{2}}\hat{c}_{0\downarrow}(\hat{c}_{-1\uparrow} + \hat{c}_{1\downarrow})(a\hat{c}_{N\uparrow}+b\hat{c}_{N\downarrow})\ket{\bold{1}}.
\end{align}

We can now quantify the particle and mode entanglement of $\ket{\psi'}$. 
Since the spin-flip transitions in Eq.~\eqref{eq:spin_flip} are a local operation on the C orbitals while the N orbitals are not affected, the mode entanglement between the C and the N orbitals should not change. Indeed, from the reduced density matrix 
\begin{equation}
\rho_N=\frac{1}{2}
\begin{pmatrix}
0&0&0&0\\
0&1&0&0\\
0&0&1&0\\
0&0&0&0\\
\end{pmatrix}
\end{equation}
we obtain an entropic mode entanglement of $S[\rho_N]=1/2$ for $\ket{\psi'}$, which is the same value as for $\ket{\psi}$.
Note that here the imposition of the particle-number superselection rule does not change the mode entanglement of $\ket{\psi'}$ since in any possible outcome of a measurement (all four lines in Eq.~\eqref{eq:NV_state_tot_after}) there are one hole in the N orbitals (\textit{Bob}) and two holes in the C orbitals (\textit{Alice}). 

The spin-flip transitions in Eq.~(\ref{eq:spin_flip}) are particle interactions which may change the particle entanglement in $\ket{\psi'}$. 
To explicitly calculate the particle entanglement of $\ket{\psi'}$, we construct its one-particle reduced density matrix in the basis  \{0$\downarrow$,1$\downarrow$,1$\uparrow$,-1$\downarrow$,-1$\uparrow$,N$\downarrow$,N$\uparrow$\} 
\begin{equation}
\rho^{(1\mathrm{p})}= \frac{1}{2}
\begin{pmatrix}
2&0&0&0&0&0&0\\
0&|b|^2&a^*b&0&0&0&0\\
0&ab^*&|a|^2&0&0&0&0\\
0&0&0&|b|^2&a^*b&0&0\\
0&0&0&ab^*&|a|^2&0&0\\
0&0&0&0&0&1&0\\
0&0&0&0&0&0&1\\
\end{pmatrix}
\end{equation}
and find an entropic particle entanglement of $S[\rho^{(1\mathrm{p})}] = 1/2[1+2(|a|^2+|b|^2)-(|a|^2+|b|^2)^2] = 1$. Surprisingly, this is the same value as for the state $\ket{\psi}$. The geometric particle entanglement stays also the same, i.e. $E_G=1/2$.  Thus, the applied spin-flip transitions do not change the particle entanglement. By taking a second look at the effect of the spin-flip interactions in Eq.~(\ref{eq:spin_flip})  we find that for $\ket{\psi}$ they simply act as a unitary spin transformation given by
\begin{align}
\hat{c}_{1\downarrow} & \rightarrow b\hat{c}_{1\downarrow}+a\hat{c}_{1\uparrow} \hspace{3.3em} 
\hat{c}_{1\uparrow}  \rightarrow -a\hat{c}_{1\downarrow}+b\hat{c}_{1\uparrow} \nonumber \\
\hat{c}_{-1\downarrow} & \rightarrow b\hat{c}_{-1\downarrow}+a\hat{c}_{-1\uparrow} \hspace{1.5em} 
\hat{c}_{-1\uparrow}  \rightarrow -a\hat{c}_{-1\downarrow}+b\hat{c}_{-1\uparrow} \nonumber \\
\hat{c}_{0\downarrow} & \rightarrow b\hat{c}_{0\downarrow}-a\hat{c}_{0\uparrow} \hspace{3.3em} 
\hat{c}_{0\uparrow}  \rightarrow a\hat{c}_{0\downarrow}+b\hat{c}_{0\uparrow}
\end{align}
which brings $\ket{\psi}$ to $\ket{\psi'}$. This explains why the particle entanglement is the same in $\ket{\psi}$ and $\ket{\psi'}$.   

The spin-flip transition localize one hole in the C 0$\downarrow$ orbital. Alice can therefore distinguish between the four different lines in Eq.~\eqref{eq:NV_state_tot_after} by measuring the occupation of the $m_l=\pm1$ spin-orbitals.
According to the outcome of this measurement she can tell Bob to apply the appropriate spin rotation and phase shift to the N orbital to obtain the original state to be teleported.

Particle entanglement plays an active role in this teleportation scheme. It is the initial particle entangled Bell state ($\ket{\psi_2}$) that allows this scheme to reach a 100\% success rate. The spin-flip transitions do not change the particle entanglement of $\ket{\psi'}$, but they allow Alice to perform an orbital occupation measurement instead of a much more involved Bell state measurement that would require a particle entangled reference state.

\subsection{Teleportation in a quantum dot array}
Semiconductor quantum dots are another promising realization of solid-state qubits for quantum computation.\cite{Loss_PRA1998,Eriksson_2004_qdot_exp} In a quantum dot, the qubit is usually represented by the spin degree of freedom of an excess electron since the decoherence time of the spin is much longer than that of the charge.\cite{Small_PRL1999} Several protocols for quantum information processing with quantum dots have already been proposed and implemented.\cite{DiVincenzo_Nature2000, Small_PRL1999, Zhan_PRL2012_hybdot} Also a few teleportation schemes based on quantum dot arrays have been suggested in Refs.~\onlinecite{Pasquale_PRL2004_qdot,Visser_PRL2006_qdot,Qiao_arxiv2019_qdot}. 
Inspired by these references, we here present as a third example a teleportation scheme involving a linear array of three quantum dots and we analyze the role of particle and mode entanglement in this scheme.

\begin{figure}[]
\includegraphics[width=6.cm]{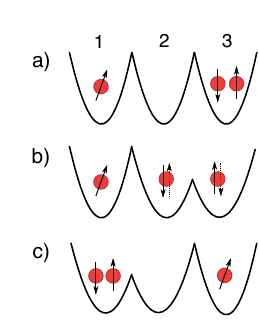}
\caption{Teleportation scheme with three quantum dots: a) The state to be teleported, Eq.~\eqref{eq:dot_tele}, is initially encoded in the spin state of an electron residing in dot 1. Two additional electrons in a singlet state are located in dot 3. b) A particle-entangled support state, Eq.~\eqref{eq:dot_supp}, is created through tunneling between dot 2 and 3. c) One possible outcome of a successful teleportation (see first line of Eq.~\eqref{eq:dot_tot_tunn_2}) obtained through a second tunneling process between dot 1 and 2. The state to be teleported is now located in dot 3.}
\label{fig:quantum_dots}
\end{figure}

In our scheme, the state to be teleported is initially encoded in the spin state of an electron in the first dot of the array (see Fig.~\ref{fig:quantum_dots}a). This state reads
\begin{equation}
\label{eq:dot_tele}
\ket{\psi_1} = (\alpha\hat{c}^\dagger_{1\uparrow} + \beta\hat{c}^\dagger_{1\downarrow})\ket{0} ,
\end{equation} 
where $\alpha$ and $\beta$ are unknown coefficients and $|\alpha|^2+|\beta|^2=1$. $\ket{\psi_1}$ is neither particle nor mode entangled, since we assign dot 1 and 2 to $Alice$ and dot 3 to $Bob$. 
Two additional electrons are residing in the third dot, i.e.    
\begin{equation}
\label{eq:dot_start}
\ket{\psi_2} = \hat{c}^\dagger_{3\uparrow}\hat{c}^\dagger_{3\downarrow}\ket{0} .
\end{equation}
These two electrons are used to produce an entangled resource state via a conditional tunneling between dot 2 and 3.  This tunneling process can be controlled by lowering and raising the tunnel barrier between the dots through a gate voltage.\cite{Loss_PRA1998} Through the tunneling, which is assisted by the local Coulomb interaction within the dots or by a total occupation measurement of dot 2 or dot 3, we can obtain the following entangled resource state (see Fig.~\ref{fig:quantum_dots}b)
\begin{equation}
\label{eq:dot_supp}
\ket{\psi_2'} = \frac{1}{\sqrt{2}}(\hat{c}^\dagger_{2\uparrow}\hat{c}^\dagger_{3\downarrow} - \hat{c}^\dagger_{2\downarrow}\hat{c}^\dagger_{3\uparrow})\ket{0}.
\end{equation} 
Similar to Eq.~\eqref{eq:NV0_init} in the second example,  $\ket{\psi_2'}$ represents a maximally particle-entangled state of two electrons with a geometric entanglement measure of $E_G=1/2$ and a particle entanglement entropy of $S=1$. 
With a bipartition between dot 2 and 3, the mode entanglement entropy of $\ket{\psi_2'}$ can easily be evaluated as $S=1/2$.

By combining $\ket{\psi_1}$ and $\ket{\psi_2'}$, we can write the total state of the quantum dot array as  
\begin{align}
\label{eq:dot_tot}
\ket{\psi}  & = \frac{1}{\sqrt{2}}(\alpha\hat{c}^\dagger_{1\uparrow} + \beta\hat{c}^\dagger_{1\downarrow})(\hat{c}^\dagger_{2\uparrow}\hat{c}^\dagger_{3\downarrow} - \hat{c}^\dagger_{2\downarrow}\hat{c}^\dagger_{3\uparrow})\ket{0} \\ \nonumber
 & =\frac{1}{\sqrt{2}}(-\alpha\hat{c}^\dagger_{1\uparrow}\hat{c}^\dagger_{2\downarrow}\hat{c}^\dagger_{3\uparrow}+\beta\hat{c}^\dagger_{1\downarrow}\hat{c}^\dagger_{2\uparrow}\hat{c}^\dagger_{3\downarrow} + \\ \nonumber
& +\alpha\hat{c}^\dagger_{1\uparrow}\hat{c}^\dagger_{2\uparrow}\hat{c}^\dagger_{3\downarrow}  - \beta\hat{c}^\dagger_{1\downarrow}\hat{c}^\dagger_{2\downarrow}\hat{c}^\dagger_{3\uparrow})\ket{0} .
\end{align}
The entanglement of $\ket{\psi}$ is the same as for $\ket{\psi_2'}$ since $\ket{\psi_1}$ does not add any entanglement.
In the next step, we allow tunneling between dot 1 and 2. In this way, we couple the state to be teleported residing in dot 1 to the entangled resource state in dots 2 and 3. First, we note that the two expressions in the last line of Eq.~\eqref{eq:dot_tot} do not allow for any tunneling  since the electrons in dot 1 and 2 have the same spin. The expressions in the second line of Eq.~\eqref{eq:dot_tot} instead allow for tunneling. Since for a 100\% success rate of this teleportation scheme we will need to have both electrons residing in either dot 1 or 2 in the end, we choose a particular conditional tunneling process given by the following conditional hopping operator which is diagonal in the bonding/anti-bonding basis defined in Eq.~(\ref{eqn:bonding}),
\begin{align}
\hat{H} & = \frac{U}{4} \hat{c}^\dagger_{a\uparrow}\hat{c}^\dagger_{a\downarrow}\hat{c}^{\phantom{\dagger}}_{a\downarrow}\hat{c}^{\phantom{\dagger}}_{a\uparrow} 
+ \frac{3U}{4} \hat{c}^\dagger_{a\downarrow}\hat{c}^\dagger_{b\uparrow}\hat{c}^{\phantom{\dagger}}_{b\uparrow}\hat{c}^{\phantom{\dagger}}_{a\downarrow}\nonumber \\
& + \frac{5U}{4} \hat{c}^\dagger_{b\uparrow}\hat{c}^\dagger_{b\downarrow}\hat{c}^{\phantom{\dagger}}_{b\downarrow}\hat{c}^{\phantom{\dagger}}_{b\uparrow} 
+ \frac{7U}{4} \hat{c}^\dagger_{a\uparrow}\hat{c}^\dagger_{b\downarrow}\hat{c}^{\phantom{\dagger}}_{b\downarrow}\hat{c}^{\phantom{\dagger}}_{a\uparrow}.
\end{align}
By transforming the creation operators in $\ket{\psi}$ to the same basis,
\begin{align}
\ket{\psi}  &=(\frac{-1}{2\sqrt{2}}\hat{c}^\dagger_{a\uparrow}\hat{c}^\dagger_{a\downarrow}+\frac{1}{2\sqrt{2}}\hat{c}^\dagger_{b\uparrow}\hat{c}^\dagger_{b\downarrow})(\alpha\hat{c}^\dagger_{3\uparrow} + \beta\hat{c}^\dagger_{3\downarrow})   \nonumber\\
&+ (\frac{1}{2\sqrt{2}}\hat{c}^\dagger_{a\downarrow}\hat{c}^\dagger_{b\uparrow} +\frac{1}{2\sqrt{2}}\hat{c}^\dagger_{a\uparrow}\hat{c}^\dagger_{b\downarrow})(\alpha\hat{c}^\dagger_{3\uparrow} - \beta\hat{c}^\dagger_{3\downarrow})  \nonumber\\
&-\frac{\alpha}{\sqrt{2}} \hat{c}^\dagger_{a\uparrow}\hat{c}^\dagger_{b\uparrow}\hat{c}^\dagger_{3\downarrow} + \frac{\beta}{\sqrt{2}} \hat{c}^\dagger_{a\downarrow}\hat{c}^\dagger_{b\downarrow}\hat{c}^\dagger_{3\uparrow}\ket{0}.
\end{align}
we can directly apply the corresponding evolution operator $W^{(2)}(t = \pi/U)$ according to Eq.~(\ref{eqn:wevolve}), giving
\begin{align}
\ket{\psi^{\prime}}  &=(\frac{-e^{\frac{i\pi}{4}}}{2\sqrt{2}}\hat{c}^\dagger_{a\uparrow}\hat{c}^\dagger_{a\downarrow}+\frac{e^{\frac{-3i\pi}{4}}}{2\sqrt{2}}\hat{c}^\dagger_{b\uparrow}\hat{c}^\dagger_{b\downarrow})(\alpha\hat{c}^\dagger_{3\uparrow} + \beta\hat{c}^\dagger_{3\downarrow})   \nonumber\\
&+ (\frac{e^{\frac{3i\pi}{4}}}{2\sqrt{2}}\hat{c}^\dagger_{a\downarrow}\hat{c}^\dagger_{b\uparrow} +\frac{e^{-\frac{i\pi}{4}}}{2\sqrt{2}}\hat{c}^\dagger_{a\uparrow}\hat{c}^\dagger_{b\downarrow})(\alpha\hat{c}^\dagger_{3\uparrow} - \beta\hat{c}^\dagger_{3\downarrow})  \nonumber\\
&-\frac{\alpha}{\sqrt{2}} \hat{c}^\dagger_{a\uparrow}\hat{c}^\dagger_{b\uparrow}\hat{c}^\dagger_{3\downarrow} + \frac{\beta}{\sqrt{2}} \hat{c}^\dagger_{a\downarrow}\hat{c}^\dagger_{b\downarrow}\hat{c}^\dagger_{3\uparrow}\ket{0}.
\end{align}
Transforming the creation operators in $\ket{\psi^{\prime}}$ back to the quantum dot basis yields   
\begin{align}
\label{eq:dot_tot_tunn}
\ket{\psi'}  & = \frac{-1}{2}[\hat{c}^\dagger_{1\uparrow} \hat{c}^\dagger_{1\downarrow}(i\alpha \hat{c}^\dagger_{3\uparrow}+\beta\hat{c}^\dagger_{3\downarrow}) 
+\hat{c}^\dagger_{2\uparrow} \hat{c}^\dagger_{2\downarrow}(\alpha\hat{c}^\dagger_{3\uparrow}+i \beta \hat{c}^\dagger_{3\downarrow})]\ket{0} \nonumber\\
& +\frac{1}{\sqrt{2}}(\alpha\hat{c}^\dagger_{1\uparrow}\hat{c}^\dagger_{2\uparrow}\hat{c}^\dagger_{3\downarrow}  - \beta\hat{c}^\dagger_{1\downarrow}\hat{c}^\dagger_{2\downarrow}\hat{c}^\dagger_{3\uparrow})\ket{0} .
\end{align}

To analyze the particle entanglement after this second tunneling process, we construct the one-particle reduced density matrix of $\ket{\psi'}$ in the basis \{1$\uparrow$,1$\downarrow$,2$\uparrow$,2$\downarrow$,3$\uparrow$,3$\downarrow$\}
\begin{align}
\rho^{(1\mathrm{p})} &= \nonumber\\ 
&\begin{pmatrix}
 \frac{1+2 \abs{\alpha}^2}{4} & 0 & 0 & -\frac{i {\beta^*} {\alpha^*}}{\sqrt{2}} & 0 & 0 \\
 0 & \frac{1+ 2 \abs{\beta}^2}{4} & 0 & 0 & 0 & 0 \\
 0 & 0 & \frac{1+2 \abs{\alpha}^2}{4} & 0 & 0 & 0 \\
 \frac{i {\beta} {\alpha}}{\sqrt{2}} & 0 & 0 & \frac{1+2 \abs{\beta}^2}{4} & 0 & 0 \\
 0 & 0 & 0 & 0 & \frac{1}{2} & 0 \\
 0 & 0 & 0 & 0 & 0 & \frac{1}{2} 
\end{pmatrix}
\end{align}
and calculate the particle entanglement entropy  $S[\rho^{(1\mathrm{p})}] = 5/4$. The geometic entanglement measure $E_G[\ket{\psi'}] = 1/2$ is obtained from a parametrized search over all Fock states. The particle entanglement entropy is larger than the one for $\ket{\psi}$, i.e. $S=1$, but the geometic entanglement measure is unchanged.
The mode entanglement does not change in the tunneling process since the tunneling Hamiltonian only acts on Alice orbitals. 
Also the particle-number superselection rule does not change the mode entanglement in this teleportation scheme since in any measurement outcome of  $\ket{\psi'}$ in Eq.~\eqref{eq:dot_tot_tunn} we have two electrons residing with \textit{Alice} (dot 1 and 2) and one with \textit{Bob} (dot 3). 

The first line of Eq.~\eqref{eq:dot_tot_tunn} already resembles  our target state for a successful teleportation, with the state to be teleported eventually residing in dot 3. In order to make use also of the second part of $\ket{\psi'}$, we need to rewrite Eq.~\eqref{eq:dot_tot_tunn} in terms of the spin in x-direction, i.e.
$\hat{c}^\dagger_{\uparrow(\downarrow)}= 1/\sqrt{2}\left(\hat{c}^\dagger_{\uparrow x}\varpm\hat{c}^\dagger_{\downarrow x}\right)$. 
This yields 
\begin{align}
\label{eq:dot_tot_tunn_2}
\ket{\psi'}  & = \frac{1}{2}\hat{c}^\dagger_{1\uparrow x} \hat{c}^\dagger_{1\downarrow x}(i\alpha \hat{c}^\dagger_{3\uparrow}+\beta\hat{c}^\dagger_{3\downarrow})\ket{0} \\ \nonumber 
& +\frac{1}{2} \hat{c}^\dagger_{2\uparrow x} \hat{c}^\dagger_{2\downarrow x}(\alpha\hat{c}^\dagger_{3\uparrow}+i\beta\hat{c}^\dagger_{3\downarrow})\ket{0} \\ \nonumber
& + \frac{1}{2\sqrt{2}}\hat{c}^\dagger_{1\uparrow x} \hat{c}^\dagger_{2\uparrow x}(\alpha\hat{c}^\dagger_{3\downarrow}-\beta\hat{c}^\dagger_{3\uparrow})\ket{0} \\ \nonumber
 & + \frac{1}{2\sqrt{2}}\hat{c}^\dagger_{1\downarrow x} \hat{c}^\dagger_{2\uparrow x}(\alpha\hat{c}^\dagger_{3\downarrow}+\beta\hat{c}^\dagger_{3\uparrow})\ket{0} \\ \nonumber
 & + \frac{1}{2\sqrt{2}}\hat{c}^\dagger_{1\uparrow x} \hat{c}^\dagger_{2\downarrow x}(\alpha\hat{c}^\dagger_{3\downarrow}+\beta\hat{c}^\dagger_{3\uparrow})\ket{0} \\ \nonumber
 & + \frac{1}{2\sqrt{2}}\hat{c}^\dagger_{1\downarrow x} \hat{c}^\dagger_{2\downarrow x}(\alpha\hat{c}^\dagger_{3\downarrow}-\beta\hat{c}^\dagger_{3\uparrow})\ket{0} ,
\end{align}
where we can now clearly identify in each line the state to be teleported --- residing in dot 3 and differing from the original state of Eq.~\eqref{eq:dot_tele} only through a unitary spin rotation. To complete the teleportation, one needs to perform a charge and/or magnetization measurement on dot 1 in order to distinguish between the different lines of Eq.~\eqref{eq:dot_tot_tunn_2}. According to the outcome of this measurement one then needs to apply a magnetic field to the electron spin in dot 3 which yields the original state to be teleported.   In Fig.~\ref{fig:quantum_dots}c)  we show one possible outcome of a successful teleportation (corresponding to the first line of Eq.~\eqref{eq:dot_tot_tunn_2}) with two electrons residing in quantum dot 1 and the state to be teleported in dot 3.

\section{Conclusion and outlook}
\label{Sec:outlook}
We have elaborated on the differences between mode and particle entanglement of electrons and shown that they both represent valuable resources for quantum information tasks such as quantum teleportation. Mode entanglement relies on a bipartition of orbitals and is closely related to the entanglement of distinguishable particles. Particle entanglement instead refers to the quantum correlations in a fermionic state which cannot be written as a Fock state (single Slater determinant). While non-local electron hopping processes in a material can lead to the formation of mode entanglement, particle interactions such as Coulomb interaction or the interaction with a detector are required to give rise to particle entanglement. 

We have  investigated the formation and the role of particle- and mode entanglement in three solid-state quantum teleportation schemes.
(i) Our first example described the teleportation of an electronic state within a hydrogen molecule on graphene. It did only require that the system starts in a mode entangled state but not a particle-entangled state. The protocol creates particle entanglement through a magnetization measurement, but this particle entanglement is shared between the system and the detector and vanish when the detector is projected out. The lack of a particle entangled state within the system itself comes at the cost of a strongly reduced success rate (50\%).
(ii) For the second teleportation scheme involving a neutral nitrogen-vacancy center NV$^0$ in diamond we started out with a highly mode and particle entangled state of two holes in the nitrogen and carbon orbitals [Eq.~\eqref{eq:NV0_init}]. The initial particle entanglement was preserved in the teleportation process until the final measurements by Alice. This protocol has a 100\% success rate in the ideal case.
(iii) In our third example --- describing a quantum teleportation in a quantum dot array --- a particle and mode entangled state [Eq.~\eqref{eq:dot_supp}] was initially created through a conditional tunneling process between two quantum dots.
Also this example has shown to yield a 100\% success rate in the ideal case.

Thus, in the investigated teleportation schemes both mode and particle entanglement are present and play an important role. Nevertheless, the state of the system (not including the detector) does not have to be particle entangled for a quantum teleportation to be successful 50\% of the times. For a 100\% teleportation success rate instead one needs to use a Bell state which is maximally particle- and mode entangled.    In our work we have pointed out the importance of particle entanglement which is often neglected in the literature.  
 In addition, our proposed electronic teleportation schemes can be a source of inspiration for the actual experimental realization of a quantum teleportation involving indistinguishable particles.

{\em Acknowledgments.} 
We thank Sumanta Bhandary and Jakob Steinbauer for fruitful discussions. A.G. acknowledges support through Schr\"odinger fellowship J-4267 of the Austrian Science Fund (FWF) and through a 'Sub auspiciis Exzellenzstipendium' of the  Austrian Federal Ministry of Education, Science and Research. PT acknowledges support from Carl Trygger's Foundation, grant number CTS 18:389.

\appendix
\section{N-SSR and particle entanglement}\label{app:nssrparticle}
In this appendix we give a short proof that if an N-electon state $\ket{\psi}$ has N-SSR restricted mode entanglement, then the state of the system after the projective orbital occupation measurement $\hat{P}^{(n)}_A \ket{\psi}$ is particle entangled for some $n$. To prove this it is enough to show that if the state $\hat{P}^{(n)}_A \ket{\psi}$ is a Fock state, then its mode entanglement is zero.

Let us start by looking at the one-particle reduce density matrix
\begin{equation}
\rho_{ij} = \bra{\psi} \hat{P}^{(n)\dagger}_A \hat{c}^{\dagger}_j \hat{c}_i \hat{P}^{(n)}_A \ket{\psi},
\end{equation}
in an orbital basis where the orbitals $i$ and $j$ belong to either partition $A$ or $B$. 
If $i \in A$ and $j \in B$ or $i \in B$ and $j \in A$  we get
\begin{equation}
\hat{P}^{(n)\dagger}_A \hat{c}^{\dagger}_j \hat{c}_i \hat{P}^{(n)}_A = 0,
\end{equation}
which imples that all off-diagonal elements in $\rho$ between the orbital partition $A$ and $B$ are zero.

If $\hat{P}^{(n)}_A \ket{\psi}$ is a Fock state then there is an orbital transformation that brings $\rho$ into a diagonal form with diagonal elements 1 or 0. Since $\rho$ has no off-diagonal elements between $A$ and $B$ this transformation can be performed without mixing the orbitals of the two partitions. In this new orbital basis we can hence write the state $\hat{P}^{(n)}_A \ket{\psi}$ as the clearly non-entangled state $\hat{S}^\dagger_\mathbf{a}\hat{S}^\dagger_\mathbf{b}\ket{0}$ where $\mathbf{a} \in \mathcal{S}^{A}_{n}$ and $\mathbf{b} \in \mathcal{S}^{B}_{N-n}$. This proves that if $\hat{P}^{(n)}_A \ket{\psi}$ has no particle entanglement, then its mode entanglement between $A$ and $B$ is also zero.

\section{Particle entanglement inequality}\label{app:entmeasure}
In this appendix we prove the particle entanglement inequality in Eq.~(\ref{eqn:entrelation}) explicitly for the linear entropy particle entanglement measure $S[\ket{\Psi}]$ and the geometric particle entanglement measure $E_G[\ket{\Psi}]$. 

The initial state in Eq.~(\ref{eqn:entrelation}) is
\begin{align}
\ket{\Psi^i} & = \sqrt{\alpha}e^{i\theta}\hat{c}^\dagger_{e}\ket{\Psi_e} + \sqrt{1-\alpha}\ket{\Psi_s}\nonumber\\
& = \Big[\sqrt{\alpha}e^{i\theta}\hat{c}^\dagger_{e}\Big(\!\!\!\!\sum_{~~\mathbf{i} \in \mathcal{S}_{N-1}}\!\!\!\!\! \Psi^e_{\mathbf{i}} \hat{S}^\dagger_{\mathbf{i}}\Big) + \sqrt{1-\alpha}\Big(\!\sum_{\mathbf{j} \in \mathcal{S}_{N}}\! \Psi^s_{\mathbf{j}} \hat{S}^\dagger_{\mathbf{j}}\Big)  \Big]\ket{0} \label{eqn:psiiapp}
\end{align}
where the states $\hat{c}^\dagger_{e}\ket{\Psi_e}$ and $\ket{\Psi_s}$ are normalized and contain N electrons. By definition neither $\ket{\Psi_e}$ nor $\ket{\Psi_s}$ contain an electron in orbital $e$. This allows us, for convenience, to absorb the relative phase $e^{i\theta}$ between into the definition of $\hat{c}^\dagger_{e}$. Futhermore, since $E[\ket{\Psi_e}] = E[\hat{c}^\dagger_{e} \ket{\Psi_e}]$ we only need to prove the inequality 
\begin{align}
E\Big[\sqrt{\alpha}\hat{c}^\dagger_{e}\ket{\Psi_e} &+ \sqrt{1-\alpha}\ket{\Psi_s}\Big] \nonumber\\
& \ge \alpha E\Big[\hat{c}^\dagger_{e}\ket{\Psi_e}\Big] + (1-\alpha)E\Big[\ket{\Psi_s}\Big].\label{eqn:entrelation3}
\end{align}

\subsection{Linear entropy measure}
The states $\hat{c}^\dagger_{e}\ket{\Psi_e}$ and $\ket{\Psi_s}$ give rise to three different contributions to the one-particle reduced density matrix, $\rho_{ij} = \bra{\Psi^i}\hat{c}^{\dagger}_j \hat{c}_i \ket{\Psi^i}$, given by
\begin{align}
\rho_{ij} &= \alpha \rho_{ij}^e + (1-\alpha) \rho_{ij}^s + \sqrt{\alpha(1-\alpha)} \rho_{ij}^{\mathrm{off}},\\
\rho_{ij}^e &= \bra{\Psi^e}\hat{c}_{e} \hat{c}^{\dagger}_j \hat{c}_i \hat{c}^\dagger_{e}\ket{\Psi^e},\\
\rho_{ij}^s &= \bra{\Psi^s}\hat{c}^{\dagger}_j \hat{c}_i \ket{\Psi^s},\\
\rho_{ij}^{\mathrm{off}} &= \bra{\Psi^e}\hat{c}_{e}\hat{c}^{\dagger}_j \hat{c}_i \ket{\Psi^s} + \bra{\Psi^s}\hat{c}^{\dagger}_j \hat{c}_i \hat{c}^\dagger_{e}\ket{\Psi^e}.\label{eqn:rhooff}
\end{align}

Let us now define the creation operator $\hat{c}^\dagger_{s}$ through the equation
\begin{equation}
\sqrt{\beta}\hat{c}^\dagger_{s}\ket{0} = (-1)^{N-1}\!\!\sum_{\mathbf{i} \in \mathcal{S}_{N-1}}\!\!\! \Psi^{e*}_{\mathbf{i}} \hat{S}_{\mathbf{i}} \ket{\Psi_s},\label{eqn:sorb}
\end{equation}
where $\beta$ is a non-negative normalization constant. Since $\hat{c}_{e} \ket{\Psi_s} = 0$ we have that $\{\hat{c}^\dagger_{s},\hat{c}_{e}\} = 0$, which implies that the orbitals $e$ and $s$ are orthogonal. By inserting $\hat{c}_{s}$ and $\hat{c}^\dagger_{s}$ into Eq.~(\ref{eqn:rhooff}) we obtain $\rho_{ij}^{\mathrm{off}} = \sqrt{\beta}(\delta_{ie}\delta_{js} + \delta_{is}\delta_{je})$. Finally, by writing $\ket{\Psi_e}$ and $\ket{\Psi_s}$ as superpositions of states that contain $\hat{c}^\dagger_{s}$ and states which do not, it is straight forward to show that $\beta \leq \rho_{ss}^s(1-\rho_{ss}^e)$.

Let us first deal with the case when $\beta = 0$, {i.e.} $\rho^{\mathrm{off}} = 0$. The linear entropy of $\rho$ is then given by
\begin{align}
S[\rho] &= \Tr\Big[ \alpha\rho^e + (1-\alpha)\rho^s - \Big(\alpha\rho^e + (1-\alpha)\rho^s\Big)^2 \Big] \nonumber\\
 &= \alpha\Tr\Big[\rho^e\Big] + (1-\alpha)\Tr\Big[\rho^s\Big] - \alpha^2\Tr\Big[(\rho^e)^2\Big] \nonumber\\
&~~ - (1-\alpha)^2\Tr\Big[(\rho^s)^2\Big] - 2\alpha(1-\alpha)\Tr\Big[\rho^e\rho^s\Big] \nonumber\\
&= \alpha\Tr\Big[\rho^e - (\rho^e)^2 \Big] + (1-\alpha)\Tr\Big[\rho^s - (\rho^s)^2\Big] \nonumber\\
&~~+ \alpha(1-\alpha)\Big(\Tr\Big[(\rho^e)^2\Big]+\Tr\Big[(\rho^e)^2\Big]-2\Tr\Big[\rho^e\rho^s\Big]  \Big).\label{eqn:rhonooff}
\end{align}
The last term in Eq.~(\ref{eqn:rhonooff}) is always larger than zero since the product of two hermitian operators $\hat{A}$ and $\hat{B}$ fulfill the Cauchy–Schwarz inequality 
\begin{align}
\Tr[\hat{A} \hat{B}] & \leq \sqrt{\Tr[\hat{A}^2]\Tr[\hat{B}^2]} \nonumber\\
& \leq \sqrt{\Tr[\hat{A}^2]\Tr[\hat{B}^2] + \frac{1}{4}\Big(\Tr[\hat{A}^2] - \Tr[\hat{B}^2]\Big)^2 } \nonumber\\
& = \frac{1}{2} (\Tr[\hat{A}^2] + \Tr[\hat{B}^2]),\label{eqn:cs}
\end{align}
which directly yeilds
\begin{align}
S[\rho] &\geq \alpha\Tr\Big[\rho^e - (\rho^e)^2 \Big] + (1-\alpha)\Tr\Big[\rho^s - (\rho^s)^2\Big] \nonumber\\
&= \alpha S[\rho^e] + (1-\alpha) S[\rho^s]\label{eqn:srzero}
\end{align}
when $\beta = 0$.

In case $\beta > 0$ we get a finite contribution from $\rho^{\mathrm{off}}$ that reduces the linear entropy of $\rho$. To prove the inequality in Eq.~(\ref{eqn:entrelation2}) it is therefore not enough to only use the Cauchy–Schwarz inequality in Eq.~(\ref{eqn:cs}) but we must treat the effect of $\rho^{\mathrm{off}}$ explicitly. To this aim, let us define the projection matrix
\begin{equation}
P_{ij} = \delta_{ei}\delta_{ej} + \delta_{si}\delta_{sj}
\end{equation}
that project out  the orbital subspace spanned by $e$ and $s$. The linear entropy of $\rho$ can then be written as 
\begin{equation}
S[\rho] = \Tr\Big[P\rho P - (P\rho P)^2 \Big] + \Tr\Big[(\rho - P\rho P) - (\rho - P\rho P)^2 \Big].\label{eqn:rhosplit}
\end{equation}
The second term in Eq.~(\ref{eqn:rhosplit}), $S[\rho - P\rho P]$, does not contain any contribution from $\rho^{\mathrm{off}}$. It can therefore be expanded as in Eq.~(\ref{eqn:rhonooff}) and bound by the inequality in Eq.~(\ref{eqn:cs}), which yields
\begin{equation}
S[\rho - P\rho P] \geq \alpha S[\rho^e - P \rho^e P] + (1-\alpha) S[\rho^s - P \rho^s P].\label{eqn:srprp}
\end{equation}
The first term in Eq.~(\ref{eqn:rhosplit}), $S[P\rho P]$, is explicitly given by
\begin{align}
S[P\rho P] & = \alpha S[P\rho^e P] + (1-\alpha)S[P\rho^s P]  \nonumber\\
&+ \alpha(1-\alpha)\Big(\Tr\Big[(P\rho^e P)^2\Big] +\Tr\Big[(P \rho^s P)^2\Big] \nonumber\\
& -2\Tr\Big[P\rho^e P \rho^s P\Big] -2\Tr\Big[\rho^{\mathrm{off}} \rho^{\mathrm{off}}\Big] \Big). \nonumber\\
&= \alpha S[P\rho^e P] + (1-\alpha)S[P\rho^s P] \nonumber\\
&~~+ \alpha(1-\alpha)\Big[1 + (\rho^{e}_{ss} - \rho^{s}_{ss})^2 -2\beta \Big]\label{eqn:sprp}
\end{align}
Substituting $\beta \leq \rho_{ss}^s(1-\rho_{ss}^e)$ into Eq.~(\ref{eqn:sprp}) yields
\begin{equation}
S[P\rho P] \geq \alpha S[P\rho^e P] + (1-\alpha)S[P\rho^s P].\label{eqn:sprpfinal}
\end{equation}

Finally, combining Eq.~(\ref{eqn:srprp}) and Eq.~(\ref{eqn:sprpfinal}) gives
\begin{equation}
S[\rho] \geq \alpha S[\rho^e] + (1-\alpha)S[\rho^s]
\end{equation}
which together with Eq.~(\ref{eqn:srzero}) proves Eq.~(\ref{eqn:entrelation3}) for the linear entropy measure for all values of $\beta$.

\subsection{Geometric measure}
Although both the geometric particle entanglement measure $E_G[\ket{\Psi}]$ and the linear entropy particle entanglement measure $S[\ket{\Psi}]$ are zero if and only if $\ket{\Psi}$ is a Fock state, they are not equivalent measures. Eq.~(\ref{eqn:entrelation3}) requires therefore a separate proof for $E_G$.

Let us start by considering an arbitrary Fock state $\ket{\Psi^{i\prime\prime}}$ with N electrons. Since $\ket{\Psi^{i\prime\prime}}$ is a Fock state there exists an orbital basis in which it can be written as a single Slater determinant $\ket{\Psi^{i\prime\prime}} = \hat{c}^{\dagger\prime\prime}_1\hat{c}^{\dagger\prime\prime}_2\cdots\hat{c}^{\dagger\prime\prime}_N \ket{0}$. We would now like to decompose $\ket{\Psi^{i\prime\prime}}$ with respect to the occupation of a given orbital $e$. To this aim, let us perform an orbital basis transformation using a unitary matrix $W$ to a basis in which $\hat{c}^{\dagger}_1 \equiv \hat{c}^{\dagger}_e$, 
\begin{equation}
\left(
\begin{array}{c}
\hat{c}^{\dagger\prime\prime}_1\\
\vdots\\
\hat{c}^{\dagger\prime\prime}_N\\
\vdots\\
\end{array}
 \right) = 
\left(
\begin{array}{cccc}
W_{11} & \cdots & W_{1N} & \cdots \\
\vdots & \ddots & \vdots \\
W_{N1} & \cdots & W_{NN} \\
\vdots & & & \ddots \\
\end{array}
\right)
\left(
\begin{array}{c}
\hat{c}^{\dagger}_1\\
\vdots\\
\hat{c}^{\dagger}_N\\
\vdots\\
\end{array}
 \right).
\end{equation}
We can always decompose the upper $N\times N$ block of $W$ into a unitary matrix $\tilde{U}$ multiplied from the right with an upper triangular matrix $\tilde{V}$ with non-negative diagonal elements $0 \leq \tilde{V}_{ii} \leq 1$. By extending $\tilde{U}$ with the identity matrix for the remaining orbitals we can define the unitary matrix $U$ that gives
%\begin{widetext}
\begin{align}
\left(
\begin{array}{cccc}
W_{11} & \cdots & W_{1N} & \cdots \\
\vdots & \ddots & \vdots \\
W_{N1} & \cdots & W_{NN} \\
\vdots & & & \ddots \\
\end{array}
\right) = 
\left(
\begin{array}{cccc}
\tilde{U}_{11} & \cdots & \tilde{U}_{1N} & 0 \\
\vdots & \ddots & \vdots & 0 \\
\tilde{U}_{N1} & \cdots & \tilde{U}_{NN} & 0\\
0 & 0 & 0 & \mathbf{1} \\
\end{array}
\right) \times & \nonumber\\
\times\left(
\begin{array}{cccc}
\tilde{V}_{11} & \cdots & \tilde{V}_{1N} & \cdots  \\
0 & \ddots & \vdots \\
0 & 0 & \tilde{V}_{NN} & \cdots \\
W_{(N+1)1} & \cdots & W_{(N+1)N} \cdots\\
\end{array}
\right)
\end{align}
%\end{widetext}
The full $V$ matrix, i.e. $V = U^{\dagger}W$, corresponds to a valid orbital basis transformation since both $U$ and $W$ are unitary. Let us therefore consider the Slater determinant given by $\ket{\Psi^{i\prime}} = \hat{c}^{\dagger\prime}_1\hat{c}^{\dagger\prime}_2\cdots\hat{c}^{\dagger\prime}_N \ket{0}$ where 
\begin{equation}
\hat{c}^{\dagger\prime}_i = \sum_{j} V_{ij} \hat{c}^{\dagger}_j.
\end{equation}
Since $V_{i1} = 0$ for $i=2,3,\cdots,N$ only $\hat{c}^{\dagger\prime}_1$ contributes to the occupation of $\hat{c}^{\dagger}_1 = \hat{c}^{\dagger}_e$ in $\ket{\Psi^{i\prime}}$. Futhermore, since $\{\hat{c}^{\dagger\prime}_1,\hat{c}^{\prime}_i\} = 0$ for $i=2,3,\cdots,N$, the same is true for the operator $\sqrt{1-V_{11}}\hat{c}_s \equiv \sum_{j=2} V_{1j} \hat{c}^{\dagger}_j$. We can therefore write~\cite{Kraus_fermion} 
\begin{equation}
\ket{\Psi^{i\prime}} = \Big(\alpha^{\prime} \hat{c}^{\dagger}_e + \sqrt{1-\alpha^{\prime}}\hat{c}^{\dagger}_s\Big)\hat{c}^{\dagger\prime}_2\cdots\hat{c}^{\dagger\prime}_N \ket{0},\label{eqn:fockdecomp}
\end{equation}
where $\alpha^{\prime} = V_{11}$. The Fock state $\ket{\Psi^{i\prime}}$ is however identical to $\ket{\Psi^{i\prime\prime}}$ since 
\begin{align}
\ket{\Psi^{i\prime}} & = \hat{c}^{\dagger\prime}_1\hat{c}^{\dagger\prime}_2\cdots\hat{c}^{\dagger\prime}_N \ket{0} \nonumber\\
& = \Big(\sum_{j=1}^N U^*_{j1} \hat{c}^{\dagger\prime\prime}_j\Big) \Big(\sum_{j=1}^N U^*_{j2} \hat{c}^{\dagger\prime\prime}_j\Big) \cdots \Big(\sum_{j=1}^N U^*_{jN} \hat{c}^{\dagger\prime\prime}_j\Big) \ket{0}\nonumber\\
& = \det{U} \hat{c}^{\dagger\prime\prime}_1\hat{c}^{\dagger\prime\prime}_2\cdots\hat{c}^{\dagger\prime\prime}_N \ket{0}\nonumber\\
& = \ket{\Psi^{i\prime\prime}}.
\end{align}
This imples that an arbitrary Fock state can always be decomposed as in Eq.~(\ref{eqn:fockdecomp}) with respect to the occupation of any single orbital $e$. 

Let us now consider the squared overlap between the Fock state $\ket{\Psi^{i\prime}}$ and the input state 
\begin{equation}
\ket{\Psi^i} = \sqrt{\alpha}\hat{c}^\dagger_{e}\ket{\Psi_e} + \sqrt{1-\alpha}\ket{\Psi_s}.
\end{equation}
The decomposition of $\ket{\Psi^{i\prime}}$ in Eq.~(\ref{eqn:fockdecomp}) gives 
\begin{align}
\abs{\braket{\Psi^{i\prime}}{\Psi^i}}^2 = &\Big|\sqrt{\alpha\alpha^{\prime}}\braket{\Psi_e^{\prime}}{\Psi_e} \nonumber\\
& + \sqrt{(1-\alpha)(1-\alpha^{\prime})}\bra{\Psi_e^{\prime}}\hat{c}_s\ket{\Psi_s}\Big|^2,
\end{align}
where $\ket{\Psi_e^{\prime}} = \hat{c}^{\dagger\prime}_2\cdots\hat{c}^{\dagger\prime}_N \ket{0}$. The overlap is maximized when the phase of $\hat{c}_s$ and the size of $\alpha^{\prime}$ is adjusted so that
\begin{align}
&\frac{\bra{\Psi_e^{\prime}}\hat{c}_s\ket{\Psi_s}}{\abs{\bra{\Psi_e^{\prime}}\hat{c}_s\ket{\Psi_s}}}  = \frac{\braket{\Psi_e^{\prime}}{\Psi_e}}{\abs{\braket{\Psi_e^{\prime}}{\Psi_e}}}, \\
&\alpha^{\prime}  = \frac{\alpha \abs{\braket{\Psi_e^{\prime}}{\Psi_e}}^2}{\alpha\abs{\braket{\Psi_e^{\prime}}{\Psi_e}}^2 + (1-\alpha)\abs{\bra{\Psi_e^{\prime}}\hat{c}_s\ket{\Psi_s}}^2},
\end{align}
which yeilds
\begin{align}
\max\limits_{\ket{\Psi^{\prime}}} & \abs{\braket{\Psi^{i\prime}}{\Psi^i}}^2 =  \nonumber\\
& = \max\limits_{\ket{\Psi_e^{\prime}},\hat{c}_s} \Big( \alpha\abs{\braket{\Psi_e^{\prime}}{\Psi_e}}^2 + (1-\alpha) \abs{\bra{\Psi_e^{\prime}}\hat{c}_s\ket{\Psi_s}}^2\Big)\nonumber\\
& \leq \alpha \max\limits_{\ket{\Psi_e^{\prime}}}\abs{\braket{\Psi_e^{\prime}}{\Psi_e}}^2  + (1-\alpha) \max\limits_{\ket{\Psi_s^{\prime}}} \abs{\bra{\Psi_s^{\prime}}\ket{\Psi_s}}^2,\label{eqn:fockmax}
\end{align}
where the maximization is over all normalized Fock states. The inequality in Eq.~(\ref{eqn:fockmax}) gives
\begin{align}
1 - \max\limits_{\ket{\Psi^{\prime}}} \abs{\braket{\Psi^{i\prime}}{\Psi^i}}^2 \geq & \alpha\Big(1 - \max\limits_{\ket{\Psi_e^{\prime}}}\abs{\braket{\Psi_e^{\prime}}{\Psi_e}}^2\Big)\nonumber\\
&+ (1-\alpha)\Big(1 - \max\limits_{\ket{\Psi_s^{\prime}}} \abs{\bra{\Psi_s^{\prime}}\ket{\Psi_s}}^2\Big),
\end{align}
which proves Eq.~(\ref{eqn:entrelation3}) for the geometric particle entanglement measure.

%\bibliography{main_bib}


\end{document}